\documentclass{article}
\usepackage{ijcai23}

\usepackage{times}
\usepackage{soul}
\usepackage{url}
\usepackage[hidelinks]{hyperref}
\usepackage[utf8]{inputenc}
\usepackage[small]{caption}
\usepackage{graphicx}
\usepackage{amsmath}
\usepackage{amsthm}
\usepackage{booktabs}
\usepackage{algorithm}
\usepackage[switch]{lineno}
\usepackage[noend]{algpseudocode}
\usepackage{natbib}
\usepackage{amssymb}
\usepackage{multirow}
\usepackage{pdfpages}

\newtheorem{example}{Example}
\newtheorem{theorem}{Theorem}
\newtheorem{lemma}{Lemma}

\title{Enhancing Datalog Reasoning with Hypertree Decompositions}

\author{
Xinyue Zhang$^1$
\and
Pan Hu$^2$\and
Yavor Nenov$^3$\And
Ian Horrocks$^1$
\affiliations
$^1$ Department of Computer Science, University of Oxford, Oxford, UK\\
$^2$ School of Electrical Information and Electronic Engineering, Shanghai Jiao Tong University, China\\
$^3$ Oxford Semantic Techonologies, Oxford, UK
\emails
\{xinyue.zhang, ian.horrocks\}@cs.ox.ac.uk,
pan.hu@sjtu.edu.cn,
yavor.nenov@oxfordsemantic.tech
}

\newcommand{\appargs}{\,\substack{\cdot\\[-0.12cm]\cdot\\[-0.12cm]\cdot\\[0.05cm]}\,}
\newcommand{\mysubsection}[1]{\smallskip\noindent\textbf{#1}:}

\newcommand{\instI}[1]{$\mathsf{inst}_{p_{#1}}^{I}$}
\newcommand{\instIPlus}[1]{$\mathsf{inst}_{p_{#1}}^{I \appargs \Delta^+}$}
\newcommand{\instIDel}[1]{$\mathsf{inst}_{p_{#1}}^{I \appargs \Delta^-}$}

\begin{document}

\maketitle

\begin{abstract}
Datalog reasoning based on the semina\"ive evaluation strategy evaluates rules using traditional join plans, which often leads to redundancy and inefficiency in practice, especially when the rules are complex. Hypertree decompositions help identify efficient query plans and reduce similar redundancy in query answering. However, it is unclear how this can be applied to materialisation and incremental reasoning with recursive Datalog programs. Moreover, hypertree decompositions require additional data structures and thus introduce nonnegligible overhead in both runtime and memory consumption. In this paper, we provide algorithms that exploit hypertree decompositions for the materialisation and incremental evaluation of Datalog programs. Furthermore, we combine this approach with standard Datalog reasoning algorithms in a modular fashion so that the overhead caused by the decompositions is reduced. 
Our empirical evaluation shows that, when the program contains complex rules, the combined approach is usually significantly faster than the baseline approach, sometimes by orders of magnitude.
\end{abstract}

\section{Introduction}
\label{sec:intro}

Datalog~\citep{abiteboul1995foundations} is a widely used rule language that can express recursive dependencies, such as graph reachability and transitive closure.
Reasoning in Datalog has found applications in different areas and supports a wide range of tasks including consistency checking~\citep{luteberget2016rule} and data analysis~\citep{alvaro2010boom}. 
Datalog is also able to capture OWL 2 RL ontologies~\citep{motik2009owl} extended with SWRL rules~\citep{horrocks2004swrl} and can thus support query answering over ontology-enriched data;
it has been implemented in a growing number of open-source and commercial systems, such as VLog~\citep{carral2019vlog}, LogicBlox~\citep{aref2015design}, Vadalog~\citep{bellomarini2018vadalog}, RDFox~\citep{nenov2015rdfox}, Oracle's database~\citep{wu2008implementing}, and GraphDB.\footnote{\url{https://graphdb.ontotext.com/}}

In a typical application, Datalog is used to declaratively represent domain knowledge as `if-then' rules. 
Given a set of explicit facts and a set of rules, Datalog systems are required to answer queries over all the facts entailed by the given rules and facts. 
To facilitate query answering, the entailed facts are often precomputed in a preprocessing step; we use \textit{materialisation} to refer to both this process and the resulting set of facts. 
Queries can then be evaluated directly over the materialisation without considering the rules.
The materialisation can be efficiently computed using the \textit{semina\"ive} algorithm~\citep{abiteboul1995foundations}, which ensures that each inference is performed only once. Incremental maintenance algorithms can then be used to avoid the cost of recomputing the materialisation when explicitly given facts are added/deleted; these include general algorithms such as the \textit{counting} algorithm~\citep{gupta1993maintaining}, the \textit{Delete/Rederive} (DRed) algorithm~\citep{staudt1995incremental}, and the \textit{Backward/Forward} (B/F) algorithm~\citep{motik2015incremental}, as well as
special purpose algorithms designed for rules with particular shapes~\citep{subercaze2016inferray}.
It has recently been shown that general and special purpose algorithms can be combined in a modular framework that supports both materialisation and incremental maintenance~\citep{hu2022modular}. 

Existing (incremental) materialisation algorithms implicitly assume that the evaluation of rule bodies is based on traditional join plans which can be suboptimal in many cases~\citep{ngo2014skew,gottlob2016hypertree}, especially in the case of cyclic rules. 
This can lead to a blow-up in the number of intermediate results and a corresponding degradation in performance (as we demonstrate in Section~\ref{sec:evaluation}).
This phenomenon can be observed in real-life applications, for example where rules are used to model complex systems, which may include the evaluation of numerical expressions.\footnote{https://2021-eu.semantics.cc/graph-based-reasoning-scaling-energy-audits-many-customers} The resulting rules are often cyclic and have large numbers of body atoms.

Similar problems also exist in query answering. One promising solution is based on \textit{hypertree decomposition}~\citep{gottlob2016hypertree}. 
Hypertree decomposition is able to decompose cyclic queries,
and Yannakakis's algorithm~\citep{yannakakis1981algorithms} can then be used to achieve efficient evaluation over the decomposition~\citep{gottlob2016hypertree}.
This method has been well-investigated with its effectiveness shown in many empirical experiments for query evaluation~\citep{tu2015duncecap,aberger2016old}.

It is unclear, however, whether the hypertree decomposition approach can benefit rule evaluation in Datalog reasoning.
Unlike query answering, which requires only a single evaluation via decomposition, rules in a Datalog program are applied multiple times until no new data can be derived. 
In this setting, it is important to avoid repetitive derivations, but this is not easy to achieve when hypertree decomposition is used for rule evaluation.
Moreover, incremental materialisation usually depends on efficiently tracking fact derivations, and it is unclear how to achieve this when such derivations depend on hypertree decomposition.
Finally, hypertree decomposition introduces some additional overhead, and this may degrade performance on simple rules.

In this paper, we introduce a Datalog reasoning algorithm that exploits hypertree decomposition to provide efficient (incremental) reasoning of recursive programs. 
Moreover, we show how this algorithm can be combined with the semina\"ive algorithm in a modular framework so as to avoid unnecessary additional overhead on simple rules.
Our empirical evaluation shows that this combined approach significantly outperforms the standard approach, sometimes by orders of magnitude, and it is never significantly slower. Our test system and data are available online.\footnote{https://xinyuezhang.xyz/HDReasoning/} Proofs and additional evaluation results are included in a technical report \citep{zhang2023enhancing}.

\section{Preliminaries}
\label{sec:pre}

\mysubsection{Datalog}
A \textit{term} is a variable or a constant. An \textit{atom} is an expression of the form $P(t_1, ..., t_k)$ where $P$ is a predicate with arity $k$, $k \geq 0$, and each $t_i$, $1 \leq i \leq k$, is a term. A \textit{fact} is a variable-free atom, and a \textit{dataset} is a finite set of facts. A \textit{rule} is an expression of the following form:
\begin{equation}
     B_0 \wedge \dots \wedge B_n \rightarrow H,
\end{equation}
where $n \geq 0$ and $B_i$, $0 \leq i \leq n$, and $H$ are atoms. For $r$ a rule, $\mathsf{h}(r) = H$ is its \textit{head}, and $\mathsf{b}(r) = \{B_0, \dots, B_n \}$ is the set of \textit{body atoms}. For $S$ an atom or a set of atoms,  $\mathsf{var}(S)$ is the set of variables appearing in $S$. For a rule $r$ to be \textit{safe}, each variable occurring in its head must also occur in at least one of its body atoms, i.e., $\mathsf{var(h}(r)) \subseteq  \mathsf{var(b}(r))$. A \textit{program} is a finite set of safe rules.

A substitution $\sigma$ is a mapping of finitely many variables to constants. For $\alpha$ a term, an atom, a rule, or a set of them, $\alpha \sigma$ is the result of replacing each occurrence of a variable $x$ in $\alpha$ with $\sigma(x)$ if $\sigma(x)$ is defined in $\sigma$. For a rule $r$ and a substitution $\sigma$, if $\sigma$ maps all the variables occurring in $r$ to constants, then $r\sigma$ is an \textit{instance} of $r$.

For a rule $r$ and a dataset $I$, 
$r[I]$ is the set of facts obtained by applying $r$ to $I$: 
\begin{equation}
\label{equ:rI}
r[I] =  \{\mathsf{h}(r\sigma)\ |\ \mathsf{b}(r\sigma) \subseteq I \}.
\end{equation}
Moreover, for a program $\Pi$ and a dataset $I$,  $\Pi[I]$ is the set obtained by applying every rule $r$ in $\Pi$ to $I$:
\begin{equation}
\Pi[I] = \bigcup_{r\in \Pi} \{ r[I] \}. 
\end{equation}

For E a dataset, let $I_0 = E$, and we define the \textit{materialisation} $I_\infty$ of $\Pi$ w.r.t. $E$ as:
\begin{equation}
    I_\infty = \bigcup_{i\geq 0} I_i,  \text{  where } I_{i}  = I_{i-1} \cup \Pi[I_{i-1}] \text{ for } i > 0.
\end{equation}
\mysubsection{Semina\"ive Algorithm}
We will briefly introduce the semina\"ive algorithm to facilitate our discussion in later sections. As we shall see, our algorithms exploit similar techniques to avoid repetition in reasoning.

\begin{algorithm}[t]
\caption{MAT($\Pi, E$)}\label{alg:semi}
\begin{algorithmic}[1]
\State $I \gets \emptyset$
\State $\Delta \gets E$
\While{$\Delta \neq \emptyset$}                                  \label{alg:semi:whileBeg}
    \State $I:=I\cup \Delta$                                   
    \State $\Delta := \Pi[I \appargs \Delta]\ \backslash\ I$     \label{alg:semi:whileEnd}
\EndWhile
\end{algorithmic}
\end{algorithm}

The semina\"ive algorithm~\citep{abiteboul1995foundations} realises non-repetitive reasoning by identifying newly derived facts in each round of rule application. Given a program $\Pi$ and a set of facts $E$, the algorithm computes the materialisation $I$ of $\Pi$ w.r.t. $E$. 
As shown in Algorithm~\ref{alg:semi}, $\Delta$ is initialised as $E$. In each round of rule applications, the algorithm will first update $I$ by adding to it the newly derived facts from the previous round and then computing 
a fresh set of derived facts using the operator $\Pi$ defined as below: 
\begin{equation}
    r[I \appargs \Delta ] = \{ \mathsf{h}(r\sigma) \ | \  \mathsf{b}(r\sigma)  \subseteq I\ \text{and}\  \mathsf{b}(r\sigma) \cap \Delta \neq \emptyset \}, \label{equ:rSemiNaive} 
\end{equation}
\begin{equation}
    \Pi[I \appargs \Delta]  = \bigcup_{r\in \Pi} \{ r[I \appargs \Delta ] \}, \label{equ:piSemiNaive}
\end{equation}
in which $\sigma$ in expression~\eqref{equ:rSemiNaive} is a substitution mapping variables in $r$ to constants, and $\Delta \subseteq I$.
The definition of $\Pi[I \appargs \Delta]$ ensures that the algorithm will only consider rule instances that have not been considered before.
In practice, $r[I \appargs \Delta]$ can be efficiently implemented by evaluating the rule body $n+1$ times \citep{motik2019maintenance}. Specifically, for the $i$th evaluation, $0 \leq i \leq n$, the body is evaluated by:
\begin{equation}
\label{exp:semiLabel}
  B_0^{I\backslash \Delta} \wedge \dots \wedge B_{i-1}^{I\backslash \Delta} \wedge B_i^{\Delta} \wedge B_{i+1}^{I} \wedge \dots \wedge B_n^{I},
\end{equation}
in which the superscript identifies the set of facts where each atom is matched.

\mysubsection{DRed Algorithm}
\label{subsec:dred}
The original DRed algorithm is presented by~\citet{gupta1993maintaining}, but it does not support non-repetitive reasoning. In this paper, we consider a non-repetitive and generalised version of the DRed algorithm presented by~\citet{hu2022modular}. This version of the DRed algorithm allows modular reasoning, i.e., reasoning over different parts of the program can be implemented using customised algorithms, which is more suitable for our discussions below.

The DRed algorithm is shown in Algorithm~\ref{alg:dred} where input arguments $\Pi$ and $E$ represent the program and the original set of explicitly given facts, $I$ is the materialisation of $\Pi$ w.r.t. $E$, and $E^+$ and $E^-$ are the sets of facts that are to be added to and deleted from $E$, respectively. As shown in lines~\ref{alg:dred:begin}--\ref{alg:dred:end}, the main idea behind DRed is to first \textit{overdelete} all possible derivations that depend on $E^-$; and then the algorithm tries to \textit{rederive} facts that have alternative proofs using the remaining facts; lastly, to \textit{add} to the materialisation, the algorithm computes the consequences of $E^+$ as well as the rederived facts. 

Specifically, overdeletion involves recursively finding all the consequences derived by $\Pi$ and  $E^-$, directly or indirectly, as shown in lines~\ref{alg:dred:overdel:begin}--\ref{alg:dred:overdel:end}. 
The function $\mathsf{Del}^r$ called in line~\ref{alg:dred:overdel:func} is intended to compute the facts that are directly affected by the deletion of $\Delta^-$. More precisely, $\mathsf{Del}^r(I, \Delta^-)$ should compute ${r[I \appargs \Delta^-]} \cap (I \backslash \Delta^-)$. Note that in line~\ref{alg:dred:overdel:func} the first argument of the call is $I \backslash D$, so it should compute ${r[I' \appargs \Delta^-]} \cap (I' \backslash \Delta^-)$ with ${I' = I \backslash D}$; the same clarification applies to $\mathsf{Red}^r$ and $\mathsf{Add}^r$, so we will not reiterate. 

The rederivation step recovers the facts that are overdeleted but are one-step provable from the remaining facts. Formally, function $\mathsf{Red}^r(I, \Delta)$ should compute ${r[I] \cap \Delta}$. Finally, during addition, the added set $N_A$ is initialised in line~\ref{alg:dred:add:initdelta}, and then from line~\ref{alg:dred:add:begin} to \ref{alg:dred:add:end} the rules are iteratively applied, similarly as in the semina\"ive algorithm. 
In this case, function $\mathsf{Add}^r(I, \Delta^+)$ is required to compute ${r[I \appargs \Delta^+]} \backslash I$.

The correctness of Algorithm~\ref{alg:dred} is guaranteed by Theorem~\ref{the:all}, which straightforwardly follows from the correctness of the modular update algorithm by~\citet{hu2022modular}.
\begin{theorem}
\label{the:all}
Algorithm~\ref{alg:dred} correctly updates the materialisation $I_\infty$ of $\Pi$ w.r.t. $E$ to $I'_\infty$ of $\Pi$ w.r.t. $E'$ where $E'=(E\backslash E^-) \cup E^+$, provided that $\mathsf{Del}^r(I, \Delta^-)$, $\mathsf{Red}^r(I, \Delta)$, and $\mathsf{Add}^r(I, \Delta^+)$ compute ${r[I \appargs \Delta^-] \cap (I \backslash \Delta^-) }$, ${r[I] \cap \Delta}$, and ${r[I \appargs \Delta^+] \backslash I}$, respectively.
\end{theorem}

Please note that the DRed algorithm could be used for the initial materialisation as well. To achieve this, we can set $E,\ I$ and $E^-$ as empty sets, and pass the set of explicitly given facts as $E^+$ to the algorithm. 

\begin{algorithm}[t]
\caption{DRed($\Pi, E, I, E^+, E^-$)}\label{alg:dred}
\begin{algorithmic}[1]
\State $D:=A:=\emptyset$, $E^- := (E^- \cap E) \backslash E^+$, $E^+ := E^+ \backslash E$
\State \Call{Overdelete}{} \label{alg:dred:begin}
\State \Call{Rederive}{}
\State \Call{Add}{} \label{alg:dred:end}
\State $E:=(E\backslash E^-) \cup E^+$, $I:=(I\backslash D) \cup A$
\vspace{0.1cm}
\Procedure{Overdelete}{}
    \State $N_D := E^-$ \label{alg:dred:overdel:begin}
    \Loop
        \State $\Delta^- := N_D \backslash D $
        \If{$\Delta^- = \emptyset$}
            \textbf{break}
        \EndIf
        \State $N_D := \emptyset$
        \For{$r \in \Pi$}
            \State $N_D := N_D \cup \mathsf{Del}^r(I\backslash D, \Delta^-)$ \label{alg:dred:overdel:func}
        \EndFor
        \State $D := D \cup \Delta^-$ \label{alg:dred:overdel:end}
    \EndLoop
\EndProcedure
\vspace{0.1cm}
\Procedure{Rederive}{}
    \State $\Delta := \emptyset$
    \For{$r \in \Pi$}
        \State $\Delta := \Delta \cup \mathsf{Red}^r(I\backslash D, D)$ \label{alg:dred:red}
    \EndFor
    \State $\Delta := \Delta \cup ((E\backslash E^-) \cap D)$
\EndProcedure
\vspace{0.1cm}
\Procedure{Add}{}
    \State $N_A := (\Delta\cup E^+) \backslash (I\backslash D)$ \label{alg:dred:add:initdelta}
    \Loop \label{alg:dred:add:begin}
        \State $\Delta^+ := N_A \backslash ((I\backslash D)\cup A)$
        \If{$\Delta^+ = \emptyset$}
            \textbf{break}
        \EndIf
        \State $A:= A\cup \Delta^+$ 
        \State $N_A := \emptyset$
        \For{$r \in \Pi$}
            \State $N_A := N_A \cup \mathsf{Add}^r((I\backslash D)\cup A, \Delta^+)$ \label{alg:dred:add:end}
        \EndFor
    \EndLoop
\EndProcedure
\end{algorithmic}
\end{algorithm}

\mysubsection{Hypertree Decomposition}
Following the definition of hypertree decomposition for conjunctive queries~\citep{gottlob2002hypertree}, we define it for Datalog rules in a similar way. 

For a Datalog rule $r$, a hypertree decomposition is a hypertree $HD = \langle T, \chi, \lambda\rangle$ in which $T=\langle N, E \rangle$ is a rooted tree, and $\chi$ associates each vertex $p\in N$ with a set of variables in $r$ whereas $\lambda$ associates $p$ with a set of atoms in $\mathsf{b}(r)$. This hypertree satisfies all the following conditions:
\begin{enumerate}
    \item for each body atom $B_i \in \mathsf{b}(r)$, there exists $p\in N$ such that $\mathsf{var}(B_i) \subseteq \chi(p);$
    \item for each variable $v \in \mathsf{var}(r)$, the set $\{p\in N\ |\ v\in \chi(p) \}$ induces a connected subtree of $T$.
    \item for each vertex $p \in N$, $\chi(p) \subseteq\mathsf{var}(\lambda(p))$.
    \item for each vertex $p \in N$, $\mathsf{var}(\lambda(p)) \cap \chi(T_p) \subseteq \chi(p)$ in which $T_p$ is the subtree of T rooted at $p$.
\end{enumerate}

The \textit{width} of the hypertree decomposition is defined as $max_{p\in N} |\lambda(p)|$. The \textit{hypertree-width hw(r)} of $r$ is the minimum width over all possible hypertree decompositions of $r$. 
In this paper, we refer to $r$ as a complex rule, or interchangeably, a cyclic rule, if and only if its hypertree width $hw(r)$ is greater than 1. 

Next, we will introduce how query evaluation works using a decomposition as join plan.
Query evaluation via hypertree decomposition is a well-investigated problem in the database literature~\citep{gottlob2016hypertree,flum2002query}, and such a process typically consists of \textit{in-node evaluation} and \textit{cross-node evaluation}. During in-node evaluation, each node $p$ in the decomposition joins the body atoms that are assigned to it (i.e., $\lambda(p)$) and stores the join results for later use. Then, cross-node evaluation applies the Yannakakis algorithm to the above join results using $T$ as the join tree. 
The standard Yannakakis algorithm in turn has two steps. The \textit{full reducer} stage applies a sequence of bottom-up left semi-joins through the tree, followed by a sequence of top-down left semi-joins using the same fixed root of the tree~\citep{bernstein1981using}. 
This removes dangling data that will not be needed in the second stage and decreases the join result size for each node. 
The \textit{cross-node join} stage joins the nodes bottom-up, and it projects to the output variables, i.e., $\mathsf{var}(\mathsf{h}(r))$, to obtain the final answers.

Overall, the (combined) complexity of query evaluation via a decomposition tree is known to be $O(v\cdot(m^k+s)\cdot log(m+s))$~\citep{gottlob2016hypertree} where $v$ is the number of variables in the query, $m$ is the cardinality of the largest relation in data, $k$ is the hypertree width of $r$, and $s$ is the output size.

\section{Motivation}
\label{sec:motivation}

\begin{figure}[t]
    \centering
    \includegraphics[width=\linewidth]{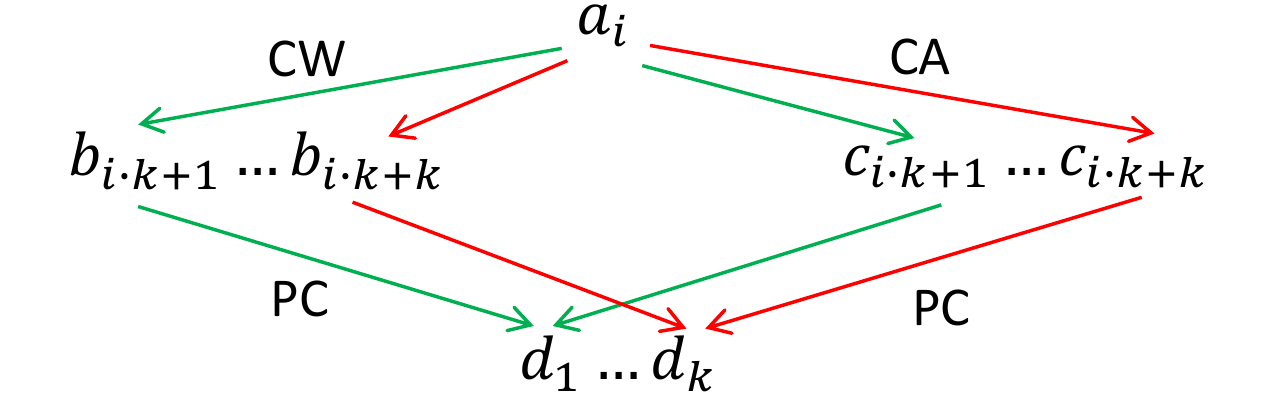}
    \caption{Illustration of joins related to a particular $a_i$, $0 \leq i < n$, in the first round of rule evaluation.}
    \label{fig:datashape}
\end{figure}

In this section, we use an example to explain how hypertree decompositions could
 benefit rule evaluation and provide some intuitions as to how they 
can be exploited in the evaluation of recursive Datalog rules. 
To this end, consider the following rule $r$, in which $\mathsf{PC, CW}$ and $\mathsf{CA}$ represent $\mathsf{PossibleCollaborator}$, $\mathsf{Coworker}$, and $\mathsf{Coauthor}$,  respectively:
\begin{equation}
    \mathsf{PC}(x,y) \leftarrow  \mathsf{CW}(x,z_1), \mathsf{CA}(x,z_2), \mathsf{PC}(z_1,y), \mathsf{PC}(z_2,y). \nonumber 
\end{equation}
Moreover, consider the dataset $E$ as specified below, where $n$ and $k$ are constants. Refer to Figure~\ref{fig:datashape} for a (partial) illustration of the dataset and the joins.
\begin{align*}
    & \{\mathsf{CW}(a_i, b_{i \cdot k + j}) \mid 0 \leq i < n, \quad 1 \leq j \leq k\}  \; \cup   \\
    & \{\mathsf{CA}(a_i, c_{i \cdot k + j}) \mid 0 \leq i < n, \quad 1 \leq j \leq k\}  \; \cup \\
    & \{\mathsf{CW}(a_n, a_2), \quad \mathsf{CA}(a_n, a_3)\} \; \cup  \\
    & \{\mathsf{PC}(b_{i \cdot k + j}, d_j) \mid 0 \leq i < n, \quad 1 \leq j \leq k\} \; \cup \\
    & \{\mathsf{PC}(c_{i \cdot k + j}, d_j) \mid 0 \leq i < n, \quad 1 \leq j \leq k\}  
\end{align*}
Each relation above contains $O(n \cdot k)$ facts, and the materialisation will 
additionally derive ${n \cdot k + k}$ facts, i.e., $\{\mathsf{PC}(a_i, d_j) \mid 0 \leq i \leq n, \quad 1 \leq j \leq k \}$. 

Now consider the first round of rule evaluation, and assume that the rule body of $r$, which corresponds to a conjunctive query, is evaluated left-to-right. Then, matching the first three atoms involves considering $O(n \cdot k^2)$ different substitutions for variables $x$, $y$, $z_1$, and $z_2$; only $O(n \cdot k)$ of them will match the last atom and eventually lead to successful derivations. In fact, one can verify that no matter how we reorder the body atoms of $r$, it will result in similar behaviour. 

Using hypertree decompositions could help process the query more efficiently. Consider decomposition $T$ of the above query consisting of two nodes $p_1$ and $p_2$, where $p_1$ is the parent node of $p_2$. Furthermore, function $\chi$ is defined as: $\chi(p_1) = \{ x, z_1, y \}$, $\chi(p_2) = \{x, z_2, y \}$, and function $\lambda$ is defined as: $\lambda(p_1) = \{\mathsf{CW}(x,z_1), \mathsf{PC}(z_1,y) \}$, $\lambda(p_2) = \{\mathsf{CA}(x,z_2), \mathsf{PC}(z_2,y) \}$. 
Recall the steps of decomposition-based query evaluation we introduced in Section~\ref{sec:pre}. During the $\textit{in-node evaluation}$ stage, each node in the decomposition will consider $O(n\cdot k)$ substitutions; the $\textit{full reducer}$ will consider $O(n\cdot k)$ substitutions and find out that nothing needs to be reduced; lastly, the \textit{cross-node evaluation} joining $p_1$ and $p_2$ also considers $O(n \cdot k)$ substitutions. Compared with the left-to-right evaluation of the query, the overall cost of this approach is $O(n \cdot k)$, as opposed to $O(n \cdot k^2)$. 
For every $a_i$ ($0 \leq i < n$), the first round of rule application will introduce additional $\mathsf{PC}$ relations between $a_i$ and $d_1$ to $d_k$ ($n\cdot k$ in total). 

Notice that rule $r$ is recursive, so the facts produced by the first round of rule evaluation could potentially lead to further derivations of the same rule. This is indeed the case in our example: the first round derives all $\mathsf{PC}(a_i, d_j)$ facts with ${0 \leq i < n}$ and ${1 \leq j \leq k}$; combined with $\{\mathsf{CW}(a_n, a_2), \mathsf{CA}(a_n, a_3)\}$ this will additionally derive ${\mathsf{PC}(a_n, d_j)}$ with ${1 \leq j \leq k}$. If we used the hypertree decomposition-based technique discussed above, then a na\"ive implementation would just add all the facts derived in the first round to the corresponding nodes and run the decomposition-based query evaluation again. However, this is unlikely to be very efficient as it would have to repeat all the work performed in the first round of rule evaluation. Ideally, we would like to make the decomposition-based query evaluation algorithm `incremental', in the sense that the algorithm minimises the amount of repeated work between different rounds of rule evaluation. As we shall see in Section~\ref{sec:alg}, this requires nontrivial adaptation of in-node evaluation, as well as the two stages of the Yannakakis algorithm. Handling incremental deletion presents another challenge, which we address following the well-known DRed algorithm.
\section{Algorithms}
\label{sec:alg}

We now introduce our reasoning algorithms based on hypertree decomposition.
We use DRed as the backbone of our algorithm,
but instead of standard reasoning algorithms with plan-based rule evaluation, we will use our hypertree decomposition-based functions $\mathsf{Del}^r$, $\mathsf{Add}^r$, and $\mathsf{Red}^r$ as discussed below. 
For each rule $r$ in $\Pi$, we assume its hypertree decomposition $\langle T^r, \chi^r, \lambda^r \rangle $ with ${T^r = \langle N^r, E^r \rangle}$ has already been computed, and $t^r$ is the root of the decomposition tree $T^r$. Our reasoning algorithms are independent of decomposition methods. 

\mysubsection{Notation}
First, analogously to expression~\eqref{equ:rSemiNaive}, for each node $p \in N^r$ we define operator $\Pi_p[I, \Delta]$, in which $I$ and $\Delta$ are sets of facts with $\Delta \subseteq I$.
\begin{equation}
\label{eq:pipidelta}
    \Pi_p[I, \Delta] = \{\chi^r(p)\sigma \ |\ \lambda^r(p) \sigma \subseteq I \text{ and } \nonumber \\
    \lambda^r(p) \sigma \cap \Delta \nsubseteq \emptyset \},
\end{equation}
Intuitively, this operator is intended to compute for a node $p$ all the instantiations influenced by the incremental update $\Delta$.
Additionally, for each node $p \in N^r$, we will make use of the following sets in the presentation of our algorithms. These sets are initialised as empty the first time DRed is executed.
\begin{enumerate}
    \item $\mathsf{inst}_p^I$ contains the join result of in-node evaluation for $p$ under the current materialisation $I$, and it is represented as tuples for variables $\chi^r(p)$. 
    Since cross-node evaluation builds upon such join results, to facilitate incremental evaluation and to avoid computing $\mathsf{inst}_p^I$ every time from scratch, $\mathsf{inst}_p^I$ has to be correctly  maintained between different executions of DRed.
    \item $\mathsf{inst}_p^{I\appargs \Delta^+}$ represents the set of instantiations that should be added to $\mathsf{inst}_p^I$ given a set of newly added facts $\Delta^+$. This set can be obtained using the operator $\Pi_p$.
    \item $\mathsf{inst}_p^{I\appargs \Delta^-}$ represents the set of instantiations that no longer hold after removing $\Delta^-$ from $I$; these instantiations should then be deleted from $\mathsf{inst}_p^I$. Similarly as above, this set can be computed using $\Pi_p$. 
    \item $\mathsf{inst}_p^{ac}$ represents the currently active instantiations that will participate in the cross-node evaluation. 
    \item $\mathsf{inst}_p^{re}$ represents the instantiations that will need to be checked during the rederivation phase.
\end{enumerate}

\begin{algorithm}[t]
\caption{$\mathsf{Add}^r[I, \Delta^+]$}\label{alg:add}
\begin{algorithmic}[1]
    \State /* in-node evaluation */
    \For{$p \in N^r$}:
        \State $\mathsf{inst}_p^{I\appargs \Delta^+} := \Pi_{p}[I, \Delta^+ ]\ \backslash\ \mathsf{inst}_p^{I}$ \label{alg:add:innode}
    \EndFor
    \State /* cross-node evaluation */
    \State $\Delta_A := \mathsf{CrossNodeEvaluation}^r(\mathsf{L^+})$ \label{alg:add:callCross}

    \State /* updating instantiations for each node */
    \For{$p \in N^r$} \label{alg:add:afterbegin}
        \State $\mathsf{inst}_{p}^{I} := \mathsf{inst}_p^{I} \cup \mathsf{inst}_p^{I \appargs \Delta^+}$ \label{alg:add:updateinsti}
        \State $\mathsf{inst}_{p}^{I \appargs \Delta^+} := \emptyset$ \label{alg:add:afterend}
    \EndFor
    \State return $\Delta_A \backslash I$ \label{alg:add:return}
\end{algorithmic}
\end{algorithm}

\mysubsection{Addition}
\label{subsec:add}
As discussed in Section~\ref{sec:motivation}, the decomposition-based query evaluation should be made incremental. To this end, Algorithm~\ref{alg:add}, which is responsible for addition, needs to distinguish between old instantiations and the new ones added due to changes in the explicitly given data. This is achieved by executing the \textit{in-node evaluation} for each node $p \in N^r$ in line~\ref{alg:add:innode} using the $\Pi_p$ operator.
Then, the \textit{cross-node evaluation} (line~\ref{alg:add:callCross})
is performed in a way similar to the evaluation of $r[I \appargs \Delta]$ outlined in Section~\ref{sec:pre}, treating each node in $T^r$ as a body atom. 
Specifically, as shown in algorithm~\ref{alg:cross}, we will evaluate the tree $|N^r|$ times. Assume that there is a fixed order among all the tree nodes for $r$, and let $p_i$, $1 \leq i \leq |N^r|$, denote the $i$th node in this ordering. Then, in the $i$th iteration of the loop of lines~\ref{alg:add:pickpivot}--\ref{alg:add:pickpivot:end}, node $p_i$ is chosen in line~\ref{alg:add:pickpivot}, and the label of each node will be 
determined by the labelling function $\mathsf{L}^+$ as specified below. In particular, node $p_i$ will be labelled $\Delta^+$; nodes preceding and succeeding $p_i$ will be labelled $I$ and $I \cup \Delta^+$, respectively.
\begin{equation}
\mathsf{L^+}(p_i, p_j) =  
\left\{
        \begin{array}{lr}
        I \cup \Delta^+, & p_j \prec p_i \\
        \Delta^+  , &  p_j = p_i \\
        I,  & p_j \succ p_i
        \end{array}.
\right.  
\end{equation}
Based on the labels assigned in line \ref{alg:add:label}, we will set $p_j^{ac}$, the active instantiations that will participate in the subsequent evaluation, as follows. Note that the last two cases will be used later for deletion.
\begin{equation}
\label{eq:initactive}
    \mathsf{inst}_p^{ac} = 
    \left\{
        \begin{array}{ll}
            \mathsf{inst}_p^I & I  \\
            \mathsf{inst}_p^{I \appargs \Delta^+} & \Delta^+ \\
            \mathsf{inst}_p^{I} \cup  \mathsf{inst}_p^{I \appargs \Delta^+}      &   I \cup \Delta^+ \\
            \mathsf{inst}_p^{I \appargs \Delta^-} & \Delta^- \\
            \mathsf{inst}_p^{I}  \backslash \mathsf{inst}_p^{I \appargs \Delta^-} & I \backslash \Delta^- 
        \end{array}.
    \right.
\end{equation}  

After fixing the active instantiations, algorithm~\ref{alg:cross} proceeds with an adapted version of the Yannakakis algorithm: lines~\ref{alg:add:fullreducer:pre}--\ref{alg:add:fullreducer} complete the \textit{full reducer} stage whereas line~\ref{alg:add:cross} performs the \textit{cross-node join}. 
By performing left semi-joins between nodes, the full reducer stage aims at deactivating instantiations that do not join and keep only the relevant ones. The standard full reducer does not consider incremental updates so adaptations are required. In particular, our incremental version of the full reducer traverses the tree three times. 
The first traversal in line~\ref{alg:add:fullreducer:pre} consists of a sequence of top-down left semi-joins with $p_i$ (the node labelled with $\Delta^+$) as the root. As $\Delta^+$ is typically smaller than the materialisation $I$, starting from $p_i$ could potentially reduce the numbers of active instantiations for the other nodes to a large extent. The second and the third traversal (line~\ref{alg:add:fullreducer}) involves applying the standard bottom-up and top-down left semi-join sequences, respectively, using the root of the decomposition tree $t^r$ as the root for the evaluation. 
Then, the cross-node join in line~\ref{alg:add:cross} evaluates the decomposition tree $T^r$ bottom-up: for each node $p \in N^r$, it joins active instantiations in $p$ with those in its children, and then projects the result to variables $\chi^r(p) \ \cup\ \mathsf{var(h}(r)) $. The join result obtained at the root $t^r$ is projected to the output variables $\mathsf{var(h}(r))$ to compute the derived facts, which are then returned to the $\mathsf{Add}^r$ function.

Lastly, lines~\ref{alg:add:afterbegin}--\ref{alg:add:afterend} of algorithm~\ref{alg:add} update the instantiations $\mathsf{inst}_p^{I}$ for each node $p$ and empty $\mathsf{inst}_p^{I \appargs \Delta^+}$ for later use.

By applying the principles of semina\"ive evaluation to both the in-node evaluation and the cross-node evaluation, 
$\mathsf{Add}^r$ avoids repeatedly reasoning over the same facts or instantiations. Lemma~\ref{the:add} states that the algorithm is correct.
\begin{lemma}
\label{the:add}
Algorithm~\ref{alg:add} computes ${r[I \appargs \Delta^+] \backslash I}$.
\end{lemma}

\begin{figure}[t]
    \centering
    \includegraphics[width=\linewidth]{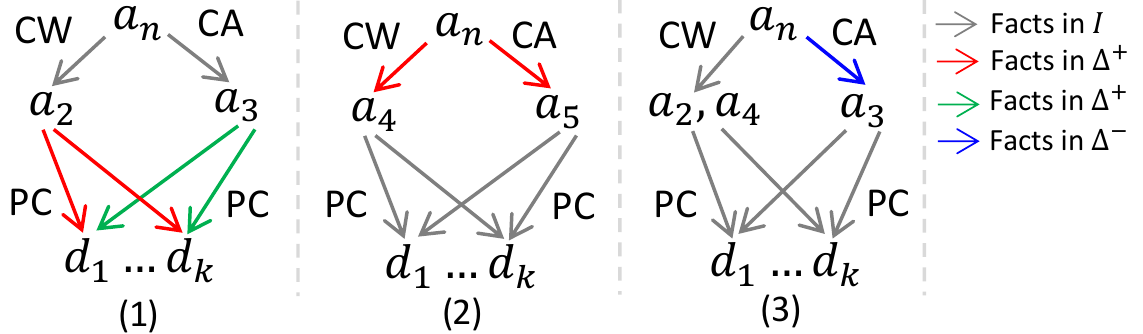}
    \caption{The illustrations depict joins in three scenarios: (1) the second round of rule evaluation; (2) the incremental rule evaluation in response to the addition of $\mathsf{CW}(a_n, a_4)$ and $\mathsf{CA}(a_n, a_5)$; and (3) the incremental rule evaluation in response to the removal of $\mathsf{CA}(a_n, a_3)$. From these joins, one can easily compute the corresponding instantiations for nodes $p_1$ and $p_2$.}
    \label{fig:joins}
\end{figure}

To further elucidate the algorithmic process, we will build upon the examples presented in Section~\ref{sec:motivation} to demonstrate our algorithm's recursive application in a step-by-step manner.
\begin{example}
Following the initial round of rule application as detailed in Section~\ref{sec:motivation}, the instantiations in $\mathsf{inst}_{p_1}^{I \appargs \Delta^+}$ and $\mathsf{inst}_{p_2}^{I \appargs \Delta^+}$ are derived in line~\ref{alg:add:innode} of algorithm~\ref{alg:add} and then merged into $\mathsf{inst}_{p_1}^{I}$ and $\mathsf{inst}_{p_2}^{I}$ in line~\ref{alg:add:updateinsti}, respectively, before being cleared in line~\ref{alg:add:afterend}. Therefore, we have ${\mathsf{inst}_{p_1}^{I} = \{(a_i, b_{i \cdot k + j}, d_j) \mid 0 \leq i < n, \; 1 \leq j \leq k \}}$, and $\mathsf{inst}_{p_2}^{I} = \{(a_i, c_{i \cdot k + j}, d_j) \mid 0 \leq i < n, \; 1 \leq j \leq k \}$. Additionally, the cross-node evaluation in line~\ref{alg:add:callCross} derives facts ${\{\mathsf{PC}(a_i, d_j) \mid 0 \leq i < n, \; 1 \leq j \leq k\}}$, which are all returned in line~\ref{alg:add:return} of the algorithm.

In the second round of application, the facts derived in the first round, i.e., $\mathsf{PC}(a_i, d_j)$ with $0 \leq i < n$ and $1 \leq j \leq k$, are passed to the $\mathsf{Add}^r$ function as $\Delta^+$. Then, line~\ref{alg:add:innode} identifies for $p_1$ the new instantiations involving facts in $\Delta^+$;  specifically, $\{(a_n, a_2, d_j) \mid 1 \leq j \leq k \}$  are assigned to $\mathsf{inst}_{p_1}^{I\appargs \Delta^+}$. Similarly, we have \instIPlus{2} $= $ $\{(a_n, a_3, d_j) \mid 1 \leq j \leq k \}$. For an illustration of the related joins, please refer to figure~\ref{fig:joins}~(1).
Then, during the cross-node evaluation, lines~\ref{alg:add:pickpivot}--\ref{alg:add:initactive} ensure that when node $p_1$ is labeled with $\Delta^+$, node $p_2$ is labeled with $I$, and so $\mathsf{inst}_{p_1}^{I\appargs \Delta^+}$ is joined with $\mathsf{inst}_{p_2}^I$, deriving no new fact. In contrast, when node $p_2$ is labeled with  $\Delta^+$, node $p_1$ is labeled with $I \cup \Delta^+$, and so \instIPlus{2} is joined with \instI{1} $\cup$ \instIPlus{1}, deriving $\mathsf{PC}(a_n, d_j)$ with $1 \leq j \leq k$. As one can readily see, the second round of rule application does not repeat work already carried out in the first round.

The above example demonstrates the process of initial materialisation. Now consider adding $\{ \mathsf{CW}(a_n, a_4)$, $\mathsf{CA}(a_n, a_5) \}$ to the explicitly given data, i.e., by setting $E^+$ to the above set of facts in the DRed algorithm. In this case, $\Delta^+$ in $\mathsf{Add}^r$ will consist of $\mathsf{CW}(a_n, a_4)$ and $\mathsf{CA}(a_n, a_5)$. Then, in line~\ref{alg:add:innode} of algorithm~\ref{alg:add}, we clearly have  $\Pi_{p_1}[I, \Delta^+] = \{(a_n, a_4, d_j) \mid 1 \leq j \leq k \}$, as illustrated by figure~\ref{fig:joins} (2). However, \instIPlus{1} will be empty since the identified instantiations already exist in \instI{1}; the same applies to $p_2$. As a result, 
no new fact is derived. This shows the benefit of keeping instantiations for the nodes of the decomposition between different runs of the DRed algorithm.
\end{example}

\begin{algorithm}[t]
\caption{$\mathsf{CrossNodeEvaluation}^r(\mathsf{L})$}\label{alg:cross}
\begin{algorithmic}[1]
        \State $\Delta := \emptyset$
       \For{$p_i \in N^r$}   /* the $\Delta$ node */ \label{alg:add:pickpivot}
        \For{$p_j \in N^r$} 
            \State $\text{label } p_j \text{ with the output of } \mathsf{L}(p_i, p_j)$  \label{alg:add:label}
            \State set $\mathsf{inst}_{p_j}^{ac}$ according to the label \label{alg:add:initactive}
        \EndFor
            \State $\mathsf{TopDownLSJ}(p_i)$ \label{alg:add:fullreducer:pre}
        \State $\mathsf{BottomUpLSJ}(t^r)$; $\mathsf{TopDownLSJ}(t^r)$ \label{alg:add:fullreducer}
        \State $\Delta := \Delta \cup \pi_{\mathsf{var(h}(r))} (\mathsf{CrossNodeJoin}(t^r))$ \label{alg:add:cross}
    \EndFor \label{alg:add:pickpivot:end}
    \State return $\Delta$
\end{algorithmic}
\end{algorithm}

\mysubsection{Deletion}
\label{subsec:overdeletion}
The $\mathsf{Del}^r$ algorithm shown in algorithm~\ref{alg:del} is analogous to $\mathsf{Add}^r$, and it identifies consequences of $r$ that are affected by the deletion of $\Delta^-$. The algorithm first computes the overdeletion $\mathsf{inst}_p^{I \appargs \Delta^- }$ using the operator $\Pi_p$ in lines~\ref{alg:del:innode:begin}--\ref{alg:del:updateInstDelta}. In addition, the instantiations that have been overdeleted are also added to $\mathsf{inst}_p^{re}$ so that they can be checked and potentially recovered during rederivation.

The \textit{cross-node evaluation} in line~\ref{alg:del:cross} is similar to that of $\mathsf{Add}^r$, except that a different labelling function $\mathsf{L}^-$ is used:
\begin{equation}
\mathsf{L}^-(p_i, p_j) =  
\left\{
        \begin{array}{lr}
        I, & p_j \prec p_i \\
        \Delta^-  , &  p_j = p_i \\
        I \backslash \Delta^-,  & p_j \succ p_i
        \end{array}.
\right.  
\end{equation}
Note that the initialisation of $\mathsf{inst}_{p_j}^{ac}$ follows equation~\eqref{eq:initactive}.
Finally, for each node $p$, the set of instantiations $\mathsf{inst}_{p}^{I}$ is updated in line~\ref{alg:del:removeInstDelta} to reflect the change, and  $\mathsf{inst}_p^{I \appargs \Delta^- }$ is emptied in line~\ref{alg:del:resetempty} for later use. 
Similarly as in $\mathsf{Add}^r$, our $\mathsf{Del}^r$ function exploits the idea of semina\"ive evaluation to avoid repeated reasoning. Lemma~\ref{the:del} states that the algorithm is correct.
\begin{lemma}
\label{the:del}
Algorithm~\ref{alg:del} computes ${r[I \appargs \Delta^-] \cap (I \backslash \Delta^-)}$.
\end{lemma}

\begin{algorithm}[t]
\caption{$\mathsf{Del}^r[I, \Delta^-]$}
\label{alg:del}
\begin{algorithmic}[1]
    \State /* in-node: overdelete */
    \For{$p \in N^r$}: \label{alg:del:innode:begin}
        \State {$\mathsf{inst}_p^{I \appargs \Delta^-} :=  \Pi_{p}[I, \Delta^- ] \cap \mathsf{inst}_p^{I}$} \label{alg:del:updateInstDelta}
        \State {$\mathsf{inst}_p^{re} := \mathsf{inst}_p^{re} \cup \mathsf{inst}_p^{I \appargs \Delta^-}$ }
    \EndFor
    
    \State /* cross-node: overdelete */
    \State $\Delta_D := \mathsf{CrossNodeEvaluation}^r(\mathsf{L^-})$ \label{alg:del:cross}

    \State /* updating instantiations for each node */
    \For{$p \in N^r$}
        \State $\mathsf{inst}_{p}^I := \mathsf{inst}_{p}^I \backslash \mathsf{inst}_{p}^{I\appargs \Delta^-}$ \label{alg:del:removeInstDelta}
        \State $\mathsf{inst}_{p}^{I \appargs \Delta^-} := \emptyset$
        \label{alg:del:resetempty}
    \EndFor
    \State return $\Delta_D \cap (I \backslash \Delta^-)$ \label{alg:del:return}
\end{algorithmic}
\end{algorithm}

The following example illustrates overdeletion using our customised algorithms.
\begin{example}
\label{exp:del}
Assume that $E^-$ is set as $\{ \mathsf{CA}(a_n, a_3) \}$ in algorithm~\ref{alg:dred}. During the overdeletion phase, $E^-$ is passed to $\mathsf{Del}^r$ as $\Delta^-$. After the execution of line~\ref{alg:del:updateInstDelta} in algorithm~\ref{alg:del}, we have \instIDel{2} $= \{(a_n, a_3, d_j) \mid 1 \leq j \leq k \}$ and \instIDel{1} $= \emptyset$, as can be seen from figure~\ref{fig:joins} (3). Then, the cross-node evaluation will derive $\mathsf{PC}(a_n, d_j)$, $1 \leq j \leq k$. These facts are temporarily overdeleted, and the rederivation stage will check whether they have alternative derivations from the remaining facts.
\end{example}

\mysubsection{Rederivation}
\label{subsec:rederivation}
The rederivation step described in algorithm~\ref{alg:red} aims at recovering facts that are overdeleted but are one-step rederivable from the remaining facts using rule $r$. In the presentation of the algorithm we take advantage of an \textit{oracle function} $\mathsf{O}$ which serves the purpose of encapsulation. The oracle function can be implemented arbitrarily, as long as it satisfies the following requirement: given a fact/tuple $f$, the oracle function returns true if $f$ has a one-step derivation from the remaining facts/tuples, and it returns false otherwise. 

In practice, there are several ways to implement such an oracle function. A straightforward way is through query evaluation. For example, to check whether a tuple $f \in \mathsf{inst}_p^{re}$ is one-step rederivable, one can construct a query using atoms in ${\lambda(p)}$, instantiate the query with the corresponding constants in $f$, and then evaluate the partially instantiated query over the remaining facts. A more advanced approach is through tracking derivation counts~\citep{hu2018optimised}: each tuple is associated with a number that indicates how many times it is derived; during reasoning, this count is incremented if a new derivation is identified, and it is decremented if a derivation no longer holds. Then, the oracle function can be realised with a simple check on the derivation count of the relevant tuple. We have adopted the second approach in this paper.

Algorithm~\ref{alg:red} proceeds as follows. First, lines~\ref{alg:red:innode:begin}--\ref{alg:red:innode} perform rederivation for in-node evaluation using the oracle. Recall that rule evaluation is decomposed into in-node evaluation and cross-node evaluation stages, so changes in the join results stored in the tree nodes have to be propagated through the decomposition tree, and this is achieved through line~\ref{alg:red:cross}. Then, lines~\ref{alg:red:final:begin}--\ref{alg:red:final:end} update the join results and clear temporal variables. Finally, line~\ref{alg:red:global} performs rederivation for cross-node evaluation and returns all the rederived facts. Lemma~\ref{the:red} states that the algorithm is correct. Together with Theorem~\ref{the:all} and Lemmas~\ref{the:add} and \ref{the:del}, this ensures the correctness of our approach.
\begin{lemma}
\label{the:red}
Algorithm~\ref{alg:red} computes ${r[I] \cap \Delta}$.
\end{lemma}

Below we continue with our running example and focus on the rederivation stage.
\begin{example}
    After the overdeletion in Example~\ref{exp:del}, we have $\mathsf{inst}_{p_2}^{re} = \{(a_n, a_3, d_j) \mid 1 \leq j \leq k \}$. These instantiations will not be recovered in line~\ref{alg:red:innode} of algorithm~\ref{alg:red} since the oracle $O$ will find out that they have no alternative derivation from the remaining data. In contrast, the overdeleted facts $\mathsf{PC}(a_n, d_j)$ with $1 \leq j \leq k$ are recovered in line~\ref{alg:red:global}. This is so since each $\mathsf{PC}(a_n, d_j)$ can be rederived using instantiation $(a_n, a_2, d_j)$ from \instI{1} and instantiation $(a_n, a_5, d_j)$ from \instI{2}.  These rederived triples are passed on to $\mathsf{Add}^r$ as $\Delta^+$, but no new fact will be derived. Overall, the removal of $\{\mathsf{CA}(a_n, a_3) \}$ do not affect the materialisation.
\end{example}

\begin{algorithm}[t]
\caption{$\mathsf{Red}^r[I, \Delta]$}
\label{alg:red}
\begin{algorithmic}[1]
    \For{$p \in N^r$} \label{alg:red:innode:begin}
        \State $\mathsf{inst}_{p}^{I \appargs \Delta^+} := \{ f \in \mathsf{inst}_p^{re}\ |\ \mathsf{O}[f] = \mathsf{true} \}$ \label{alg:red:innode}
    \EndFor
    \State $\Delta_R := \mathsf{CrossNodeEvaluation}^r(\mathsf{L^+}) \cap \Delta $ \label{alg:red:cross}
    \For{$p \in N^r$} \label{alg:red:final:begin}
        \State $\mathsf{inst}_{p}^{I} := \mathsf{inst}_p^{I} \cup \mathsf{inst}_p^{I \appargs \Delta^+}$ \label{alg:red:updatinginstI}
        \State $\mathsf{inst}_p^{re} := \mathsf{inst}_{p}^{I \appargs \Delta^+} := \emptyset$ 
        \label{alg:red:final:end}
    \EndFor
    \State return $\Delta_R \cup \{ f \in \Delta\ |\ \mathsf{O}[f] = \mathsf{true} \}$ \label{alg:red:global}
\end{algorithmic}
\end{algorithm}

\section{Implementation and Evaluation}
\label{sec:evaluation}

\begin{table}[t]
\centering
\resizebox{\columnwidth}{!}{
\begin{tabular}{c|ccccc}
\toprule
Benchmarks   & $|E|$ & $|I|$ & $|\Pi|$ & $|\Pi_s|$ & $|\Pi_c|$ \\ 
\midrule
LUBM L   & 66,751,196 & 91,128,727  & 98 & 98 & 0  \\
LUBM L+C & 66,751,196 & 99,361,809 & 114 & 98 & 16 \\
Exp      & 3,362,280  & 6,440,280   & 3  & 0 & 3  \\ 
YAGO     & 58,276,868 & 59,755,990  & 23 & 0 & 23 \\
\bottomrule
\end{tabular}
}
\caption{Dataset statistics. $|\Pi_s|$ and $|\Pi_c|$ refer to the numbers of simple and complex rules, respectively.}
\label{tab:data}
\end{table}

\begin{table*}[t]
\centering
{\footnotesize
\resizebox{\textwidth}{!}{
\begin{tabular}{c|rrrr|rrrr|rrrr|rrrr}
\toprule
\multirow{2}{*}{Method} & \multicolumn{4}{c|}{materialisation}  & \multicolumn{4}{c|}{small deletions} & \multicolumn{4}{c|}{large deletions} & \multicolumn{4}{c}{small additions} \\ 
         &  \multicolumn{1}{c}{L+C} &  \multicolumn{1}{c}{L} & \multicolumn{1}{c}{Exp}  & \multicolumn{1}{c|}{YAGO}   &  \multicolumn{1}{c}{L+C}  &  \multicolumn{1}{c}{L} & \multicolumn{1}{c}{Exp} & \multicolumn{1}{c|}{YAGO}  &  \multicolumn{1}{c}{L+C}  &  \multicolumn{1}{c}{L} & \multicolumn{1}{c}{Exp} & \multicolumn{1}{c|}{YAGO}   &  \multicolumn{1}{c}{L+C}  &  \multicolumn{1}{c}{L} & \multicolumn{1}{c}{Exp} & \multicolumn{1}{c}{YAGO} \\ 
         \midrule
standard & 29,577.90 & 95.73 & 7,039.87 & 155,022.00  & 0.92 & 0.03 & 37.60 &  20.06 & 15,193.70 & 27.09  & 4,006.44 & 126,562.00 & 0.97            & 0.02    & 40.23 &  20.42 \\
HD       & 1,168.83       & 740.81 & 56.83  & 367.59  & 4.00           & 3.70    & 0.47  & 0.18 & 812.32        & 558.90 & 30.93  & 168.34 & 1.04           & 0.45    & 0.57 & 0.17 \\
combined & 554.00      & 75.50       & 57.01   &  366.03   & 1.06          & 0.04       & 0.45 &  0.20    & 195.51         & 21.71      & 28.62  &   159.43   & 0.73         & 0.06     & 0.53 & 0.17 \\ 
\bottomrule
\end{tabular}
}}
\caption{Materialisation and incremental reasoning time in seconds}
\label{tab:expriments}
\end{table*}

To evaluate our algorithms we have developed a proof-of-concept implementation and conducted several experiments.

\mysubsection{Implementation}
The algorithms presented in Section~\ref{sec:alg} are independent of the choice of decompositions; however, different hypertree decompositions will lead to very different performance even if they share the same hypertree width. This is because the decomposition method only considers structural information of the queries and ignores quantitative information of the data. To address this problem, \citet{scarcello2007weighted} introduced an algorithm that chooses the optimal decomposition w.r.t.\ a given cost model. 
We adopt this algorithm with a cost model consisting of two parts: (1) an estimate of the cost of intra-node evaluation, i.e., the joins among $\lambda(p)$; and (2) an estimate of the cost of inter-node evaluation, i.e., the joins between nodes. In our implementation, for (1), we use the standard textbook cardinality estimator described in Chapter 16.4 of the book~\citep{garcia2008database} to estimate the cardinality of $\underset{B_i \in \lambda(p)}{\large\bowtie} B_i$ for a node $p$; for (2), we use $2*(|\hat{p_i}|+|\hat{p_j}|)$ to estimate the cost of performing semi-joins between nodes $p_i$ and $p_j$, where $|\hat{p_i}|$ and $|\hat{p_j}|$ represent the estimated node size.

Moreover, the extra step of full reducer we introduced in algorithm~\ref{alg:cross} (line~\ref{alg:add:fullreducer:pre}) is more suitable for small updates, in which the node with the smallest size helps reduce other large nodes. If the size of all the nodes is comparable, then this step would be unnecessary. Therefore, in practice, we only perform this optimisation if the number of active instantiations in the $\Delta$ node (i.e., $p_i$) is more than three times smaller than the maximum number of active instantiations in each node.

\mysubsection{Benchmarks}
We tested our system using the well-known LUBM and YAGO benchmarks~\citep{guo2005lubm,suchanek2008yago}, and a synthetic Exp (\textit{expressions}) benchmark which we created to capture complex rule patterns that commonly occur in practice. 
LUBM models a university domain, and it includes a data generator that can create datasets of varying sizes;
we used the LUBM-500 dataset which includes data for 500 universities.
Since the ontology of LUBM is not in OWL 2 RL, we use the LUBM~L variant created by~\citet{zhou2013making}. 
The LUBM~L rules are very simple, so we added 16 rules that capture more complex but semantically reasonable relations in the domain; some of these rules are rewritten from the cyclic queries used by~\citet{stefanoni2018estimating};
we call the resulting rule set LUBM~L+C. 
One example rule is $\mathsf{SA}(?p_1, ?p_2) \leftarrow \mathsf{HA}(?o. ?p_1), \mathsf{AD}(?p_1, ?ad), \mathsf{HA}(?o, ?p_2), \mathsf{AD}(?p_2, ?ad)$, 
in which $\mathsf{HA}$ and $\mathsf{AD}$ represent predicates $\mathsf{hasAlumnus}$ and $\mathsf{hasAdvisor}$ respectively, while $\mathsf{SA}$ in the head represents a new predicate $\mathsf{haveSameAdvisor}$ that links pairs of students $?p_1$ and $?p_2$ (not necessarily distinct) who are from the same university $?o$ and share the same advisor $?ad$. 

YAGO is a real-world RDF dataset with general knowledge about movies, people, organisations, cities, and countries. We rewrote 23 cyclic queries with different topologies (i.e., cycle, clique, petal, flower) used by~\citet{park2020g} into 19 non-recursive rules and 4 recursive rules. These rules are helpful to evaluate the performance of our algorithm on topologies that are frequently observed in real-world graph queries~\citep{bonifati2017analytical}.

As mentioned in Section~\ref{sec:intro}, realistic applications often involve complex rules. 
One example is the use of rules to evaluate numerical expressions, and our Exp benchmark has been created to simulate such cases. 
Specifically, Exp applies Datalog rules to evaluate expression trees of various depths. It contains three recursive rules capturing the arithmetical operations \textit{addition}, \textit{subtraction} and \textit{multiplication}; 
each of these rules is cyclic and contains 9 body atoms.
A generator is used to create data for a given number of expressions, sets of values and maximum depth. In our evaluation we generated 300 expressions, each with 300 values and a maximum depth of 5.
Details of the three benchmarks are given in Table~\ref{tab:data}, where $|E|$ is the number of given facts, $|I|$ is the number of facts in the materialisation, and $|\Pi|$, $|\Pi_s|$, $|\Pi_c|$ are the numbers of rules, simple rules, and complex rules, respectively.

\mysubsection{Compared Approaches}
We considered three different approaches. The \textit{standard} approach uses the semina\"ive algorithm for materialisation and an optimised variant of DRed for incremental maintenance. 
The \textit{HD} approach uses our hypertree decomposition based algorithms. The \textit{combined}  
approach applies HD algorithms to complex rules and standard algorithms to the remaining rules. 
To ensure fairness, all three approaches are implemented on top of the same code base obtained from the authors of the modular reasoning framework
~\citep{hu2022modular}. The framework allows us to partition a program into modules and apply custom algorithms to each module as required.

\mysubsection{Test Setups}
All of our experiments are conducted on a Dell PowerEdge R730 server with 512GB RAM and 2 Intel Xeon E5-2640 2.60GHz processors, running Fedora 33, kernel version 5.10.8. We evaluate the running time of \textit{materialisation} (the initial materialisation with all the explicitly given facts inserted as $E^+$ in Algorithm~\ref{alg:dred}), \textit{small deletions} (randomly selecting 1,000 facts from the dataset as $E^-$ in Algorithm~\ref{alg:dred}), \textit{large deletions} (randomly selecting 25\% of the dataset as $E^-$), and \textit{small additions} (adding 1,000 facts as $E^+$ into the dataset). Materialisation can be regarded as a \textit{large addition}.


\mysubsection{Analysis}
The experimental results are shown in Table~\ref{tab:expriments} in which L and L+C are short for LUBM L and LUBM L+C respectively. The computation of decompositions takes place during initial materialisation only and the time taken is included in the materialisation time reported in Table 2; it takes less than 0.05 seconds in all cases.
As can be seen, 
the combined approach outperforms the other approaches in most cases, sometimes by a large factor, and it is slower than the standard approach only for some of the small update tasks on LUBM L and L+C where processing time is generally small.
In contrast, the standard approach performs poorly when complex rules are included (i.e., L+C, YAGO, and Exp), while the HD approach performs poorly on the simple rules in LUBM L. 
In particular, our combined approach is 75-139x faster than the standard approach for all the tasks on Exp; on YAGO, it is 100-793x faster. Moreover, for the materialisation and large deletion tasks on LUBM L+C, the combined approach is about 53x and 77x faster than the standard approach, respectively.
Furthermore, for the small deletion and addition tasks on LUBM L+C and all the tasks on LUBM L, our combined method achieves a comparable result with the standard approach.
The combined approach performs similarly to the standard approach on LUBM L, as the HD module is never invoked (there are no cyclic rules), and it performs similarly to the HD approach on Exp and YAGO, as the HD module is always invoked (all rules are cyclic).
Our evaluation illustrates the benefit of the hypertree decomposition-based algorithms when processing complex rules,  and it shows that by combining HD algorithms with standard reasoning algorithms in a modular framework we can enjoy this benefit without degrading performance in cases where some or all of the rules are relatively simple.

Finally, the HD algorithms have to maintain auxiliary data structures for rule evaluation, which incurs some space overhead when the HD module is invoked. Specifically, our combined method consumes up to 2.3 times the memory consumed by the standard algorithm; the detailed memory consumption for each setting can be found in the technical report~\citep{zhang2023enhancing}.

\section{Related Work}
\mysubsection{HD in Query Answering}
The HD methods have been used in database systems to optimise the performance of query answering. 
For RDF workload, \citet{aberger2016old} evaluated empirically the use of HD combined with worst-case optimal join algorithms, showing up to 6x performance advantage on bottleneck cyclic RDF queries. 
Also, in the EmptyHeaded~\citep{aberger2017emptyheaded} relational engine, a query compiler has been implemented to choose the order of attributes in multiway joins based on a decomposition. This line of work focuses on optimising the evaluation of a single query, while our work focuses on evaluating recursive Datalog rules. 
For a more comprehensive review of HD techniques for query answering, please refer to~\cite{gottlob2016hypertree}.

\mysubsection{HD in Answer Set Programming}
\citet{jakl2009answer} applied HD techniques to the evaluation of propositional answer set programs. Assuming that the treewidth of a program is fixed as a constant, they devise fixed-parameter tractable algorithms for key ASP problems including consistency checking, counting the number of answer sets of a program, and enumerating such answers. In contrast to our work, their research focuses on propositional answer set programs.

For ASP in the non-ground setting, a program is usually grounded first, and then a solver deals with the ground instances. The usage of (hyper)tree decomposition has been investigated to decrease the size of generated ground rules in the grounding phase~\citep{bichler2020lpopt,calimeri2019optimizing}. \citet{bichler2020lpopt} used hypertree decomposition as a guide to rewrite a larger rule into several smaller rules, thus 
reducing the number of considered groundings;  \citet{calimeri2019optimizing} studied several heuristics that could predict in advance whether a decomposition is beneficial. 
In contrast, our work focuses on the (incremental) evaluation directly over the decomposition since the decomposition solely cannot avoid the potential blowup during the evaluation of the smaller rules.

\section{Perspectives}

In this paper, we introduced a hypertree decomposition-based reasoning algorithm, which supports rule evaluation, incremental reasoning, and recursive reasoning. 
We implemented our algorithm in a modular framework such that the overhead caused by using decomposition is incurred only for complex rules, and demonstrate empirically that this approach is effective on simple, complex and mixed rule sets.

Despite the promising results, we see many opportunities for further improving the performance of the presented algorithms. Firstly, our decomposition remains unchanged once it is fixed. However, as the input data and the materialisation change over time, the initial decomposition may no longer be optimal for rule evaluation. It would be beneficial if the maintenance could be done with the underlying decomposition changing. However, this would be challenging since the data structure in each decomposition node is maintained based on the previous decomposition, and changing the decomposition would require transferring information from the old node to the new one.

Secondly, although the memory usage has been optimised to some extent, intermediate results still take up a significant amount of space. This problem could be mitigated by incrementally computing the final join result without explicitly storing the intermediate results, or by storing only ``useful" intermediate results. 

Finally, it would be interesting to adapt our work to Datalog extensions, such as Datalog$^\pm$~\citep{cali2011datalog+} and DatalogMTL~\citep{walega2019datalogmtl}.
This would require introducing mechanisms to process the relevant additional features, such as the existential quantifier in Datalog$^\pm$ and the use of intervals in DatalogMTL.

\section*{Acknowledgements}
This work was supported by the following EPSRC projects: OASIS (EP/S032347/1), UK FIRES (EP/S019111/1), and ConCur (EP/V050869/1), as well as by SIRIUS Center for Scalable Data Access, Samsung Research UK, and NSFC grant No. 62206169.

\bibliographystyle{named}
\bibliography{ijcai23}

\begin{thebibliography}{}

\bibitem[\protect\citeauthoryear{Aberger \bgroup \em et al.\egroup
  }{2016}]{aberger2016old}
Christopher~R Aberger, Susan Tu, Kunle Olukotun, and Christopher R{\'e}.
\newblock Old techniques for new join algorithms: A case study in rdf
  processing.
\newblock In {\em 2016 IEEE 32nd International Conference on Data Engineering
  Workshops (ICDEW)}, pages 97--102. IEEE, 2016.

\bibitem[\protect\citeauthoryear{Aberger \bgroup \em et al.\egroup
  }{2017}]{aberger2017emptyheaded}
Christopher~R Aberger, Andrew Lamb, Susan Tu, Andres N{\"o}tzli, Kunle
  Olukotun, and Christopher R{\'e}.
\newblock Emptyheaded: A relational engine for graph processing.
\newblock {\em ACM Transactions on Database Systems (TODS)}, 42(4):1--44, 2017.

\bibitem[\protect\citeauthoryear{Abiteboul \bgroup \em et al.\egroup
  }{1995}]{abiteboul1995foundations}
Serge Abiteboul, Richard Hull, and Victor Vianu.
\newblock {\em Foundations of databases}, volume~8.
\newblock Addison-Wesley Reading, 1995.

\bibitem[\protect\citeauthoryear{Alvaro \bgroup \em et al.\egroup
  }{2010}]{alvaro2010boom}
Peter Alvaro, Tyson Condie, Neil Conway, Khaled Elmeleegy, Joseph~M
  Hellerstein, and Russell Sears.
\newblock Boom analytics: exploring data-centric, declarative programming for
  the cloud.
\newblock In {\em Proceedings of the 5th European conference on Computer
  systems}, pages 223--236, 2010.

\bibitem[\protect\citeauthoryear{Aref \bgroup \em et al.\egroup
  }{2015}]{aref2015design}
Molham Aref, Balder ten Cate, Todd~J Green, Benny Kimelfeld, Dan Olteanu, Emir
  Pasalic, Todd~L Veldhuizen, and Geoffrey Washburn.
\newblock Design and implementation of the logicblox system.
\newblock In {\em Proceedings of the 2015 ACM SIGMOD International Conference
  on Management of Data}, pages 1371--1382, 2015.

\bibitem[\protect\citeauthoryear{Bellomarini \bgroup \em et al.\egroup
  }{2018}]{bellomarini2018vadalog}
Luigi Bellomarini, Georg Gottlob, and Emanuel Sallinger.
\newblock The vadalog system: Datalog-based reasoning for knowledge graphs.
\newblock {\em arXiv preprint arXiv:1807.08709}, 2018.

\bibitem[\protect\citeauthoryear{Bernstein and Chiu}{1981}]{bernstein1981using}
Philip~A Bernstein and Dah-Ming~W Chiu.
\newblock Using semi-joins to solve relational queries.
\newblock {\em Journal of the ACM (JACM)}, 28(1):25--40, 1981.

\bibitem[\protect\citeauthoryear{Bichler \bgroup \em et al.\egroup
  }{2020}]{bichler2020lpopt}
Manuel Bichler, Michael Morak, and Stefan Woltran.
\newblock lpopt: A rule optimization tool for answer set programming.
\newblock {\em Fundamenta Informaticae}, 177(3-4):275--296, 2020.

\bibitem[\protect\citeauthoryear{Bonifati \bgroup \em et al.\egroup
  }{2017}]{bonifati2017analytical}
Angela Bonifati, Wim Martens, and Thomas Timm.
\newblock An analytical study of large sparql query logs.
\newblock {\em arXiv preprint arXiv:1708.00363}, 2017.

\bibitem[\protect\citeauthoryear{Cal{\`\i} \bgroup \em et al.\egroup
  }{2011}]{cali2011datalog+}
Andrea Cal{\`\i}, Georg Gottlob, Thomas Lukasiewicz, and Andreas Pieris.
\newblock Datalog+/-: A family of languages for ontology querying.
\newblock In {\em Datalog Reloaded: First International Workshop, Datalog 2010,
  Oxford, UK, March 16-19, 2010. Revised Selected Papers}, pages 351--368.
  Springer, 2011.

\bibitem[\protect\citeauthoryear{Calimeri \bgroup \em et al.\egroup
  }{2019}]{calimeri2019optimizing}
Francesco Calimeri, Simona Perri, and Jessica Zangari.
\newblock Optimizing answer set computation via heuristic-based decomposition.
\newblock {\em Theory and Practice of Logic Programming}, 19(4):603--628, 2019.

\bibitem[\protect\citeauthoryear{Carral \bgroup \em et al.\egroup
  }{2019}]{carral2019vlog}
David Carral, Irina Dragoste, Larry Gonz{\'a}lez, Ceriel Jacobs, Markus
  Kr{\"o}tzsch, and Jacopo Urbani.
\newblock Vlog: A rule engine for knowledge graphs.
\newblock In {\em International Semantic Web Conference}, pages 19--35.
  Springer, 2019.

\bibitem[\protect\citeauthoryear{Flum \bgroup \em et al.\egroup
  }{2002}]{flum2002query}
J{\"o}rg Flum, Markus Frick, and Martin Grohe.
\newblock Query evaluation via tree-decompositions.
\newblock {\em Journal of the ACM (JACM)}, 49(6):716--752, 2002.

\bibitem[\protect\citeauthoryear{Garcia-Molina}{2008}]{garcia2008database}
Hector Garcia-Molina.
\newblock {\em Database systems: the complete book}.
\newblock Pearson Education India, 2008.

\bibitem[\protect\citeauthoryear{Gottlob \bgroup \em et al.\egroup
  }{2002}]{gottlob2002hypertree}
Georg Gottlob, Nicola Leone, and Francesco Scarcello.
\newblock Hypertree decompositions and tractable queries.
\newblock {\em Journal of Computer and System Sciences}, 64(3):579--627, 2002.

\bibitem[\protect\citeauthoryear{Gottlob \bgroup \em et al.\egroup
  }{2016}]{gottlob2016hypertree}
Georg Gottlob, Gianluigi Greco, Nicola Leone, and Francesco Scarcello.
\newblock Hypertree decompositions: Questions and answers.
\newblock In {\em Proceedings of the 35th ACM SIGMOD-SIGACT-SIGAI Symposium on
  Principles of Database Systems}, pages 57--74, 2016.

\bibitem[\protect\citeauthoryear{Guo \bgroup \em et al.\egroup
  }{2005}]{guo2005lubm}
Yuanbo Guo, Zhengxiang Pan, and Jeff Heflin.
\newblock Lubm: A benchmark for owl knowledge base systems.
\newblock {\em Journal of Web Semantics}, 3(2-3):158--182, 2005.

\bibitem[\protect\citeauthoryear{Gupta \bgroup \em et al.\egroup
  }{1993}]{gupta1993maintaining}
Ashish Gupta, Inderpal~Singh Mumick, and Venkatramanan~Siva Subrahmanian.
\newblock Maintaining views incrementally.
\newblock {\em ACM SIGMOD Record}, 22(2):157--166, 1993.

\bibitem[\protect\citeauthoryear{Horrocks \bgroup \em et al.\egroup
  }{2004}]{horrocks2004swrl}
Ian Horrocks, Peter~F Patel-Schneider, Harold Boley, Said Tabet, Benjamin
  Grosof, Mike Dean, et~al.
\newblock Swrl: A semantic web rule language combining owl and ruleml.
\newblock {\em W3C Member submission}, 21(79):1--31, 2004.

\bibitem[\protect\citeauthoryear{Hu \bgroup \em et al.\egroup
  }{2018}]{hu2018optimised}
Pan Hu, Boris Motik, and Ian Horrocks.
\newblock Optimised maintenance of datalog materialisations.
\newblock In {\em Proceedings of the AAAI Conference on Artificial
  Intelligence}, volume~32, 2018.

\bibitem[\protect\citeauthoryear{Hu \bgroup \em et al.\egroup
  }{2022}]{hu2022modular}
Pan Hu, Boris Motik, and Ian Horrocks.
\newblock Modular materialisation of datalog programs.
\newblock {\em Artificial Intelligence}, 308:103726, 2022.

\bibitem[\protect\citeauthoryear{Jakl \bgroup \em et al.\egroup
  }{2009}]{jakl2009answer}
Michael Jakl, Reinhard Pichler, and Stefan Woltran.
\newblock Answer-set programming with bounded treewidth.
\newblock In {\em IJCAI}, volume~9, pages 816--822, 2009.

\bibitem[\protect\citeauthoryear{Luteberget \bgroup \em et al.\egroup
  }{2016}]{luteberget2016rule}
Bj{\o}rnar Luteberget, Christian Johansen, and Martin Steffen.
\newblock Rule-based consistency checking of railway infrastructure designs.
\newblock In {\em International Conference on Integrated Formal Methods}, pages
  491--507. Springer, 2016.

\bibitem[\protect\citeauthoryear{Motik \bgroup \em et al.\egroup
  }{2009}]{motik2009owl}
Boris Motik, Peter~F Patel-Schneider, Bijan Parsia, Conrad Bock, Achille
  Fokoue, Peter Haase, Rinke Hoekstra, Ian Horrocks, Alan Ruttenberg, Uli
  Sattler, et~al.
\newblock Owl 2 web ontology language: Structural specification and
  functional-style syntax.
\newblock {\em W3C recommendation}, 27(65):159, 2009.

\bibitem[\protect\citeauthoryear{Motik \bgroup \em et al.\egroup
  }{2015}]{motik2015incremental}
Boris Motik, Yavor Nenov, Robert Piro, and Ian Horrocks.
\newblock Incremental update of datalog materialisation: the backward/forward
  algorithm.
\newblock In {\em Proceedings of the AAAI Conference on Artificial
  Intelligence}, volume~29, 2015.

\bibitem[\protect\citeauthoryear{Motik \bgroup \em et al.\egroup
  }{2019}]{motik2019maintenance}
Boris Motik, Yavor Nenov, Robert Piro, and Ian Horrocks.
\newblock Maintenance of datalog materialisations revisited.
\newblock {\em Artificial Intelligence}, 269:76--136, 2019.

\bibitem[\protect\citeauthoryear{Nenov \bgroup \em et al.\egroup
  }{2015}]{nenov2015rdfox}
Yavor Nenov, Robert Piro, Boris Motik, Ian Horrocks, Zhe Wu, and Jay Banerjee.
\newblock Rdfox: A highly-scalable rdf store.
\newblock In {\em International Semantic Web Conference}, pages 3--20.
  Springer, 2015.

\bibitem[\protect\citeauthoryear{Ngo \bgroup \em et al.\egroup
  }{2014}]{ngo2014skew}
Hung~Q Ngo, Christopher R{\'e}, and Atri Rudra.
\newblock Skew strikes back: New developments in the theory of join algorithms.
\newblock {\em ACM SIGMOD Record}, 42(4):5--16, 2014.

\bibitem[\protect\citeauthoryear{Park \bgroup \em et al.\egroup
  }{2020}]{park2020g}
Yeonsu Park, Seongyun Ko, Sourav~S Bhowmick, Kyoungmin Kim, Kijae Hong, and
  Wook-Shin Han.
\newblock G-care: A framework for performance benchmarking of cardinality
  estimation techniques for subgraph matching.
\newblock In {\em Proceedings of the 2020 ACM SIGMOD International Conference
  on Management of Data}, pages 1099--1114, 2020.

\bibitem[\protect\citeauthoryear{Scarcello \bgroup \em et al.\egroup
  }{2007}]{scarcello2007weighted}
Francesco Scarcello, Gianluigi Greco, and Nicola Leone.
\newblock Weighted hypertree decompositions and optimal query plans.
\newblock {\em Journal of Computer and System Sciences}, 73(3):475--506, 2007.

\bibitem[\protect\citeauthoryear{Staudt and
  Jarke}{1995}]{staudt1995incremental}
Martin Staudt and Matthias Jarke.
\newblock {\em Incremental maintenance of externally materialized views}.
\newblock Citeseer, 1995.

\bibitem[\protect\citeauthoryear{Stefanoni \bgroup \em et al.\egroup
  }{2018}]{stefanoni2018estimating}
Giorgio Stefanoni, Boris Motik, and Egor~V Kostylev.
\newblock Estimating the cardinality of conjunctive queries over rdf data using
  graph summarisation.
\newblock In {\em Proceedings of the 2018 World Wide Web Conference}, pages
  1043--1052, 2018.

\bibitem[\protect\citeauthoryear{Subercaze \bgroup \em et al.\egroup
  }{2016}]{subercaze2016inferray}
Julien Subercaze, Christophe Gravier, Jules Chevalier, and Frederique Laforest.
\newblock Inferray: fast in-memory rdf inference.
\newblock In {\em VLDB}, volume~9, 2016.

\bibitem[\protect\citeauthoryear{Suchanek \bgroup \em et al.\egroup
  }{2008}]{suchanek2008yago}
Fabian~M Suchanek, Gjergji Kasneci, and Gerhard Weikum.
\newblock Yago: A large ontology from wikipedia and wordnet.
\newblock {\em Journal of Web Semantics}, 6(3):203--217, 2008.

\bibitem[\protect\citeauthoryear{Tu and R{\'e}}{2015}]{tu2015duncecap}
Susan Tu and Christopher R{\'e}.
\newblock Duncecap: Query plans using generalized hypertree decompositions.
\newblock In {\em Proceedings of the 2015 ACM SIGMOD International Conference
  on Management of Data}, pages 2077--2078, 2015.

\bibitem[\protect\citeauthoryear{Walega \bgroup \em et al.\egroup
  }{2019}]{walega2019datalogmtl}
Przemyslaw~Andrzej Walega, B~Cuenca~Grau, Mark Kaminski, and Egor~V Kostylev.
\newblock Datalogmtl: Computational complexity and expressive power.
\newblock International Joint Conferences on Artificial Intelligence, 2019.

\bibitem[\protect\citeauthoryear{Wu \bgroup \em et al.\egroup
  }{2008}]{wu2008implementing}
Zhe Wu, George Eadon, Souripriya Das, Eugene~Inseok Chong, Vladimir Kolovski,
  Melliyal Annamalai, and Jagannathan Srinivasan.
\newblock Implementing an inference engine for rdfs/owl constructs and
  user-defined rules in oracle.
\newblock In {\em 2008 IEEE 24th International Conference on Data Engineering},
  pages 1239--1248. IEEE, 2008.

\bibitem[\protect\citeauthoryear{Yannakakis}{1981}]{yannakakis1981algorithms}
Mihalis Yannakakis.
\newblock Algorithms for acyclic database schemes.
\newblock In {\em VLDB}, volume~81, pages 82--94, 1981.

\bibitem[\protect\citeauthoryear{Zhang \bgroup \em et al.\egroup
  }{2023}]{zhang2023enhancing}
Xinyue Zhang, Pan Hu, Yavor Nenov, and Ian Horrocks.
\newblock Enhancing datalog reasoning with hypertree decompositions.
\newblock {\em CoRR}, abs/2305.06854, 2023.

\bibitem[\protect\citeauthoryear{Zhou \bgroup \em et al.\egroup
  }{2013}]{zhou2013making}
Yujiao Zhou, Bernardo Cuenca~Grau, Ian Horrocks, Zhe Wu, and Jay Banerjee.
\newblock Making the most of your triple store: query answering in owl 2 using
  an rl reasoner.
\newblock In {\em Proceedings of the 22nd international conference on World
  Wide Web}, pages 1569--1580, 2013.

\end{thebibliography}



\section*{Supplementary material}

# Enhancing Datalog Reasoning with Hypertree Decompositions - Appendix

## A  Memory consumption

We measured both the peak memory usage during the materialisation and the static memory usage consumed by the triple tables and indexes used in standard approaches, as well as auxiliary data structures used in HD modules. Table 1 and 2 show the detailed peak and static memory usage, respectively. And the last line of both tables illustrates the ratio of space consumed by the combined method to that consumed by the standard approach.

| Benchmarks | L | L+C | Exp | YAGO |
|---|---|---|---|---|
| standard | 7.2 | 8.4 | 0.6 | 5.2 |
| HD | 20.1 | 29.4 | 1.2 | 9.5 |
| combined | 7.2 | 16.5 | 1.2 | 9.5 |
| times | 1.00 | 1.96 | 2.25 | 1.85 |

Table 1: Peak memory usage (in GB)

| Benchmarks | L | L+C | Exp | YAGO |
|---|---|---|---|---|
| standard | 5.5 | 6.3 | 0.5 | 3.8 |
| HD | 18.7 | 28.1 | 1.1 | 4.8 |
| combined | 5.5 | 14.6 | 1.1 | 4.8 |
| times | 1.00 | 2.30 | 1.77 | 1.27 |

Table 2: Static memory usage (in GB)

## B  Algorithms

**Algorithm 1** DRed($\Pi, E, I, E^+, E^-$)

1: $D := A := \emptyset$, $E^- := (E^- \cap E)\setminus E^+$, $E^+ := E^+\setminus E$
2: OVERDELETE
3: REDERIVE
4: ADD
5: $E := (E\setminus E^-) \cup E^+$, $I := (I\setminus D) \cup A$
6: **procedure** OVERDELETE
7:     $N_D := E^-$
8:     **loop**
9:         $\Delta^- := N_D\setminus D$
10:        **if** $\Delta^- = \emptyset$ **then break**
11:        $N_D := \emptyset$
12:        **for** $r \in \Pi$ **do**
13:            $N_D := N_D \cup \text{Del}^r(I\setminus D, \Delta^-)$
14:        $D := D \cup \Delta^-$
15: **procedure** REDERIVE
16:    $\Delta := \emptyset$
17:    **for** $r \in \Pi$ **do**
18:        $\Delta := \Delta \cup \text{Red}^r(I\setminus D, D)$
19:    $\Delta := \Delta \cup ((E\setminus E^-) \cap D)$
20: **procedure** ADD
21:    $N_A := (\Delta \cup E^+)\setminus(I\setminus D)$
22:    **loop**
23:        $\Delta^+ := N_A\setminus((I\setminus D) \cup A)$
24:        **if** $\Delta^+ = \emptyset$ **then break**
25:        $A := A \cup \Delta^+$
26:        $N_A := \emptyset$
27:        **for** $r \in \Pi$ **do**
28:            $N_A := N_A \cup \text{Add}^r((I\setminus D) \cup A, \Delta^+)$

**Algorithm 2** $\text{Add}^r[I, \Delta^+]$

1: /* in-node evaluation */
2: **for** $p \in N^r$ **do**:
3:     $\text{inst}_p^{I:\Delta^+} := \Pi_p[I, \Delta^+] \setminus \text{inst}_p^I$
4: /* cross-node evaluation */
5: $\Delta_A := \text{CrossNodeEvaluation}^r(\text{L}^+)$
6: /* updating instantiations for each node */
7: **for** $p \in N^r$ **do**
8:     $\text{inst}_p^I := \text{inst}_p^I \cup \text{inst}_p^{I:\Delta^+}$
9:     $\text{inst}_p^{I:\Delta^+} := \emptyset$
10: **return** $\Delta_A \setminus I$

---

**Algorithm 3** $\text{CrossNodeEvaluation}^r(\text{L})$

1: $\Delta := \emptyset$
2: **for** $p_i \in N^r$ **do** /* the $\Delta$ node */
3:     **for** $p_j \in N^r$ **do**
4:         label $p_j$ with the output of $\text{L}(p_i, p_j)$
5:         set $\text{inst}_{p_j}^{ac}$ according to the label
6:     $\text{TopDownLSJ}(p_i)$
7:     $\text{BottomUpLSJ}(t^r); \text{TopDownLSJ}(t^r)$
8:     $\Delta := \Delta \cup \pi_{\text{var}(h(r))}(\text{CrossNodeJoin}(t^r))$
9: **return** $\Delta$

## C   Proof of some claims

Before providing the proofs for the three lemmas, we first prove several useful claims, which will facilitate our discussion later[1]. In particular, we will show that during the execution of functions $\text{Add}^r$, $\text{Del}^r$, and $\text{Red}^r$ made in the DRed algorithm, the intermediate result $\text{inst}_p^I$ of each node $p \in N^r$ for every $r$ is correctly maintained.

To better capture the computation of $\text{inst}_p^I$, we define $\Pi_p[I]$ as:

$$\Pi_p[I] = \{\chi^r(p)\sigma \mid \lambda^r(p)\sigma \subseteq I\}, \tag{1}$$

in which $\sigma$ is a substitution. Recall that $\Pi_p[I, \Delta]$ is already defined in section 4 of the main submission, as follows:

$$\Pi_p[I, \Delta] = \{\chi^r(p)\sigma \mid \lambda^r(p)\sigma \subseteq I \text{ and } \\ \lambda^r(p)\sigma \cap \Delta \nsubseteq \emptyset\}, \tag{2}$$

in which $I$ and $\Delta$ are sets of facts with $\Delta \subseteq I$.

Also, since $\text{inst}_p^{I:\Delta^+}$ is computed by $\Pi_p[I, \Delta]$ in line 3 of algorithm 2, we could safely infer that $\text{inst}_p^{I:\Delta^+}$ should contain the tuples as below:

$$\text{inst}_p^{I:\Delta^+} = \{\chi^r(p)\sigma \mid \lambda^r(p)\sigma \nsubseteq I \setminus \Delta^+ \text{ and } \\ \lambda^r(p)\sigma \subseteq I \text{ and } \chi^r(p)\sigma \notin \text{inst}_p^I\}, \tag{3}$$

in which $\Delta^+$ is newly added tuples and $\Delta^+ \subseteq I$, $\sigma$ is every possible substitution defined on $\text{var}(\lambda^r(p))$ (similarly below).

---
[1]The reference to algorithms in the following sections are based on the algorithm stated in appendix B unless specified.

---

**Algorithm 4** $\text{Del}^r[I, \Delta^-]$

1: /* in-node: overdelete */
2: **for** $p \in N^r$ **do**:
3:     $\text{inst}_p^{I:\Delta^-} := \Pi_p[I, \Delta^-] \cap \text{inst}_p^I$
4:     $\text{inst}_p^{re} := \text{inst}_p^{re} \cup \text{inst}_p^{I:\Delta^-}$
5: /* cross-node: overdelete */
6: $\Delta_D := \text{CrossNodeEvaluation}^r(\text{L}^-)$
7: /* updating instantiations for each node */
8: **for** $p \in N^r$ **do**
9:     $\text{inst}_p^I := \text{inst}_p^I \setminus \text{inst}_p^{I:\Delta^-}$
10:     $\text{inst}_p^{I:\Delta^-} := \emptyset$
11: **return** $\Delta_D \cap (I \setminus \Delta^-)$

---

**Algorithm 5** $\text{Red}^r[I, \Delta]$

1: **for** $p \in N^r$ **do**
2:     $\text{inst}_p^{I:\Delta^+} := \{f \in \text{inst}_p^{re} \mid O[f] = \text{true}\}$
3: $\Delta_R := \text{CrossNodeEvaluation}^r(\text{L}^+) \cap \Delta$
4: **for** $p \in N^r$ **do**
5:     $\text{inst}_p^I := \text{inst}_p^I \cup \text{inst}_p^{I:\Delta^+}$
6:     $\text{inst}_p^{re} := \text{inst}_p^{I:\Delta^+} := \emptyset$
7: **return** $\Delta_R \cup \{f \in \Delta \mid O[f] = \text{true}\}$

---

Similarly, for $\text{inst}_p^{I:\Delta^-}$ that is also computed by $\Pi_p[I, \Delta]$ in line 3 of algorithm 4, we have:

$$\text{inst}_p^{I:\Delta^-} = \{\chi^r(p)\sigma \mid \lambda^r(p)\sigma \nsubseteq I \setminus \Delta^- \text{ and } \\ \lambda^r(p)\sigma \subseteq I \text{ and } \chi^r(p)\sigma \in \text{inst}_p^I\}, \tag{4}$$

in which $\Delta^-$ is newly deleted tuples and $\Delta^- \subseteq I$.

By expressions (3) and (4) above, we could notice that the values of $\text{inst}_p^{I:\Delta^+}$ and $\text{inst}_p^{I:\Delta^-}$ also depend on the previous status of $\text{inst}_p^I$. Moreover, $\text{inst}_p^I$ will be updated by $\text{inst}_p^{I:\Delta^+}$ in line 8 of $\text{Add}^r$, line 5 of $\text{Red}^r$ and by $\text{inst}_p^{I:\Delta^-}$ in line 9 of $\text{Del}^r$. Therefore, whether $\text{inst}_p^I$ is correctly maintained is related to whether $\text{inst}_p^I$ was correctly maintained before.

Observe that during the execution of DRed, function calls to $\text{Add}^r$, $\text{Del}^r$, and $\text{Red}^r$ are not made in an arbitrary order, but follow a certain pattern. As such, following Hu *et al.* [2022], we define a call history $H^r$ of the form (5) for each rule $r$ as a finite and nonempty sequence of *runs* with length $m$; each $Q_i^r$ with $0 \leq i \leq m$ is a finite and nonempty sequence of *calls* of the form (6); each $C_{i,j}^r$ is a call of function $\text{Add}^r$, $\text{Del}^r$, or $\text{Red}^r$.

$$H^r = Q_0^r, \ldots, Q_m^r, \tag{5}$$
$$Q_i^r = C_{i,1}^r, \ldots, C_{i,h_i}^r. \tag{6}$$

Intuitively, each run represents a sequence of calls for $r$ during one execution of the DRed algorithm. For example, $Q_0^r$ represents the initial materialisation, and thus it would be one

Red$^r$ call, followed by a series of Add$^r$ calls (note that when $E = \emptyset$ function Del$^r$ is not called due to line 10). Moreover, for each $0 \leq i \leq m$ and $1 \leq j \leq h_i$, let $(I^r_{i,j-1}, \Delta^r_{i,j-1})$ be the arguments passed to call $C^r_{i,j}$; additionally, for each $1 \leq i \leq m$, let $I^r_{i-1,h_{i-1}} = I^r_{i,0}$.

We first show for each $r$, inst$^I_p$ is correctly maintained in the call sequence of $Q^r_0$. Since for the first time that Red$^r$ is called, both its parameters are empty, so we start our proof by discussing Add$^r$, i.e., $C^r_{0,j}$ for $1 < j \leq h_0$.

**Claim 1.** *For every node $p$, every call of* Add$^r[I, \Delta^+]$ *($C^r_{0,j}(I^r_{0,j-1}, \Delta^r_{0,j-1})$ with $1 < j \leq h_0$) can correctly update the set* inst$^I_p$ *from* $\Pi_p[I^r_{0,j-1}\backslash\Delta^r_{0,j-1}]$ *to* $\Pi_p[I^r_{0,j-1}]$ *during the initial materialisation ($Q^r_0$).*

Since Add$^r$ is called recursively in $Q^r_0$ and the correctness of claim 1 also depends on whether inst$^I_p$ is updated correctly when Add$^r$ is called last time, therefore, we use induction to prove claim 1.

1. *Induction Base*: If the current call is $C^r_{0,2}(I^r_{0,1}, \Delta^r_{0,1})$, then before $C^r_{0,2}(I^r_{0,1}, \Delta^r_{0,1})$ is executed, inst$^I_p$ is (initialised as) an empty set, which satisfies $\Pi_p[I^r_{0,1}\backslash\Delta^r_{0,1}] = \Pi_p[\emptyset] = \emptyset$.

2. *Induction Steps*: The current call of Add$^r$ ($C^r_{0,j}(I^r_{0,j-1}, \Delta^r_{0,j-1})$ with $1 < j \leq h_0$) can correctly update inst$^I_p$ if: (1) the values of inst$^I_p$ before updating is $\Pi_p[I^r_{0,j-1}\backslash\Delta^r_{0,j-1}]$; (2) in line 3 of algorithm 2, inst$_p^{I:\Delta^+}$ can be correctly updated to $\Pi_p[I^r_{0,j-1}]\backslash\Pi_p[I^r_{0,j-1}\backslash\Delta^r_{0,j-1}]$, then inst$^I_p$ can be correctly updated to $\Pi_p[I^r_{0,j-1}]$ in line 8 by adding inst$_p^{I:\Delta^+}$ to inst$^I_p$.

For (1), if Add$^r$ is called for the first time (i.e. $C^r_{0,2}$) or the previous call of Add$^r$ ($C^r_{0,j-1}(I^r_{0,j-2}, \Delta^r_{0,j-2})$ when $2 < j \leq h_0$) updates inst$^I_p$ correctly (i.e., to $\Pi_p[I^r_{0,j-2}] = \Pi_p[I^r_{0,j-1}\backslash\Delta^r_{0,j-1}]$), then (1) is satisfied. For (2), we will prove it by showing its soundness and completeness.

(2a) **Soundness**: for every $f = \chi^r(p)\sigma$ that $f \notin \Pi_p[I^r_{0,j-1}]\backslash\Pi_p[I^r_{0,j-1}\backslash\Delta^r_{0,j-1}]$ in which $\sigma$ is the corresponding substitution, when line 3 of Add$^r$ is executed $f \notin \Pi_p[I^r_{0,j-1}, \Delta^r_{0,j-1}]\backslash$inst$^I_p$ (or equivalently, $f \notin$ inst$_p^{I:\Delta^+}$).

(i) If $f \in \Pi_p[I^r_{0,j-1}]$ and $f \in \Pi_p[I^r_{0,j-1}\backslash\Delta^r_{0,j-1}]$, then $f \in$ inst$^I_p$, therefore, $f \notin \Pi_p[I^r_{0,j-1}, \Delta^r_{0,j-1}]\backslash$inst$^I_p$; (ii) If $f \notin \Pi_p[I^r_{0,j-1}]$, then $\lambda^r(p)\sigma \nsubseteq I^r_{0,j-1}$. According to the expression (2) of $\Pi_p[I, \Delta]$, $f \notin \Pi_p[I^r_{0,j-1}, \Delta^r_{0,j-1}]$. Therefore, $f \notin \Pi_p[I^r_{0,j-1}, \Delta^r_{0,j-1}]\backslash$inst$^I_p$.

(2b) **Completeness**: for every $f = \chi^r(p)\sigma$ that $f \in \Pi_p[I^r_{0,j-1}]$ but $f \notin \Pi_p[I^r_{0,j-1}\backslash\Delta^r_{0,j-1}]$, $f$ can be derived by line 3 and inserted to inst$_p^{I:\Delta^+}$.

Since $f \in \Pi_p[I^r_{0,j-1}]$ but $f \notin \Pi_p[I^r_{0,j-1}\backslash\Delta^r_{0,j-1}]$, we can have $\lambda^r(p)\sigma \subseteq I^r_{0,j-1}$ but $\lambda^r(p)\sigma \nsubseteq I^r_{0,j-1}\backslash\Delta^r_{0,j-1}$. Equivalently, $\lambda^r(p)\sigma \cap \Delta^r_{0,j-1} \neq \emptyset$. According to the expression (2) of $\Pi_p[I, \Delta]$, $f \in \Pi_p[I^r_{0,j-1}, \Delta^r_{0,j-1}]$. Because of the condition (1) above, when line 3 is executed, inst$^I_p$ equals to $\Pi_p[I^r_{0,j-1}\backslash\Delta^r_{0,j-1}]$. Therefore, $f \in \Pi_p[I^r_{0,j-1}, \Delta^r_{0,j-1}]\backslash$inst$^I_p$ such that $f \in$ inst$_p^{I:\Delta^+}$. $\square$

Without loss of generality, we can assume for each run $Q^r_i$ when $i \geq 1$, it consists of a series of calls of Del$^r$ ($C^r_{i,1}$ to $C^r_{i,d_i}$), one call of Red$^r$ ($C^r_{i,d_i+1}$), and a series of calls of Add$^r$ ($C^r_{i,d_i+2}$ to $C^r_{i,h_i}$) where $d_i$ is the number of deletion calls. We will first show this series Del$^r$ calls in $Q^r_1$ will maintain the inst$^I_p$ correctly.

**Claim 2.** *For the first incremental updates (i.e., run $Q^r_1$), the first call of* Del$^r$ *(i.e., $C^r_{1,1}[I^r_{1,0}, \Delta^r_{1,0}]$) can update* inst$^I_p$ *correctly from* $\Pi_p[I^r_{1,0}]$ *to* $\Pi_p[I^r_{1,0}\backslash\Delta^r_{1,0}]\backslash\Pi_p[I^r_{1,0}, \Delta^r_{1,0}]$.

We first show after $Q^r_0$ and before any calls of $Q^r_1$, the values in inst$^I_p$ is exactly $\Pi_p[I^r_{1,0}]$. As we defined above, $I^r_{1,0} = I^r_{0,h_0}$. Following claim 1, we can have after $C^r_{0,h_0}$, inst$^I_p$ can be updated to $\Pi_p[I^r_{0,h_0}] = \Pi_p[I^r_{1,0}]$.

Then, we can prove that in line 3, $\Pi_p[I^r_{1,0}]\backslash(\Pi_p[I^r_{1,0}\backslash\Delta^r_{1,0}]\backslash\Pi_p[I^r_{1,0}, \Delta^r_{1,0}])$ ($F^r_{1,0}$ for simplicity) will be added to inst$_p^{I:\Delta^-}$ and then removed from inst$^I_p$ in line 9.

1. **Soundness**: For every $f = \chi^r(p)\sigma \notin F^r_{1,0}$, $f$ will not be inserted to inst$_p^{I:\Delta^-}$.

   There are two cases:
   (1a) $f \notin \Pi_p[I^r_{1,0}]$ and $f \notin \Pi_p[I^r_{1,0}, \Delta^r_{1,0}]$.
   (1b) $f \in \Pi_p[I^r_{1,0}]$, $f \in \Pi_p[I^r_{1,0}\backslash\Delta^r_{1,0}]$, but $f \notin \Pi_p[I^r_{1,0}, \Delta^r_{1,0}]$.
   Both cases consist of $f \notin \Pi_p[I^r_{1,0}, \Delta^r_{1,0}]$, therefore, there is no way $f$ will be included in inst$_p^{I:\Delta^-}$ in line 3.

2. **Completeness**: For every $f = \chi^r(p)\sigma \in F^r_{1,0}$, $f$ will be inserted to inst$_p^{I:\Delta^-}$.

   Since $f \in F^r_{1,0}$, then there are several cases: $f \in \Pi_p[I^r_{1,0}]$, and (1) $f \notin \Pi_p[I^r_{1,0}\backslash\Delta^r_{1,0}]$; (2) $f \in \Pi_p[I^r_{1,0}\backslash\Delta^r_{1,0}]$, but $f \in \Pi_p[I^r_{1,0}, \Delta^r_{1,0}]$. For (1), we have $\lambda^r(p)\sigma \subseteq I^r_{1,0}$, and $\lambda^r(p)\sigma \nsubseteq I^r_{1,0}\backslash\Delta^r_{1,0}$. According to the expression (2), $f \in \Pi_p[I^r_{1,0}, \Delta^r_{1,0}]$, then $f$ will be added to inst$_p^{I:\Delta^-}$ in line 3. For (2), we have $\lambda^r(p)\sigma \subseteq I^r_{1,0}$, $\lambda^r(p)\sigma \subseteq I^r_{1,0}\backslash\Delta^r_{1,0}$, but $f \in \Pi_p[I^r_{1,0}, \Delta^r_{1,0}]$. In this case, since $f \in \Pi_p[I^r_{1,0}, \Delta^r_{1,0}]$, there exists another $\sigma'$ s.t. $\chi^r(p)\sigma = \chi^r(p)\sigma'$, $\lambda^r(p)\sigma' \subseteq I^r_{1,0}$, and $\lambda^r(p)\sigma' \nsubseteq I^r_{1,0}\backslash\Delta^r_{1,0}$. Therefore, $f$ will be overdeleted in line 3. $\square$

For simplicity, we first define a symbol that will be used below:
$$L_{i,j} := \Pi_p[I^r_{i,j+1}]\backslash\Pi_p[I^r_{i,j}, \Delta^r_{i,j}].$$

**Claim 3.** *For the first incremental updates (i.e., run $Q^r_1$), the calls of* Del$^r$ *after the first call (i.e.,*

$C^r_{1,j+1}[I^r_{1,j}, \Delta^r_{1,j}]$ for $1 \leq j \leq d_1$) can update $\mathsf{inst}^I_p$ correctly from $\Pi_p[I^r_{1,j}]\backslash\Pi_p[I^r_{1,j-1}, \Delta^r_{1,j-1}]$ ($L_{1,j-1}$) to $\Pi_p[I^r_{1,j}\backslash\Delta^r_{1,j}]\backslash\Pi_p[I^r_{1,j}, \Delta^r_{1,j}]$ (since $I^r_{1,j}\backslash\Delta^r_{1,j} = I^r_{1,j+1}$, so $\Pi_p[I^r_{1,j}\backslash\Delta^r_{1,j}]\backslash\Pi_p[I^r_{1,j}, \Delta^r_{1,j}]$ can be represented by $L_{1,j}$).

1. *Induction Base*: Before $C^r_{1,2}$, the values of $\mathsf{inst}^I_p$ is exactly $L_{1,j-1}$ (i.e., $L_{1,0}$). Following claim 2, $\mathsf{inst}^I_p$ is $\Pi_p[I^r_{1,0}\backslash\Delta^r_{1,0}]\backslash\Pi_p[I^r_{1,0}, \Delta^r_{1,0}]$, and $I^r_{1,1} = I^r_{1,0}\backslash\Delta^r_{1,0}$, so $\mathsf{inst}^I_p$ satisfy $\Pi_p[I^r_{1,j}]\backslash\Pi_p[I^r_{1,j-1}, \Delta^r_{1,j-1}]$ when $j = 1$.

2. *Induction Step*: The current call of $\mathsf{Del}^r$ ($C^r_{1,j+1}(I^r_{1,j}, \Delta^r_{1,j})$ with $2 \leq j \leq d_1$) can correctly update $\mathsf{inst}^I_p$ if: (1) the values of $\mathsf{inst}^I_p$ before updating is $L_{1,j-1}$; (2) in line 3 of algorithm 4, $\mathsf{inst}^{I:\Delta^-}_p$ can be correctly updated to $L_{1,j-1}\backslash L_{1,j}$, then $\mathsf{inst}^I_p$ can be correctly updated to $L_{1,j}$ in line 9 by removing $\mathsf{inst}^{I:\Delta^-}_p$ from $\mathsf{inst}^I_p$.

For (1), if it is the call $C^r_{1,2}$ or the previous $\mathsf{Del}^r$ calls $C^r_{1,j}$ with $2 < j \leq d_1$ update $\mathsf{inst}^I_p$ correctly, then (1) is satisfied. For (2), we will prove it by showing its soundness and completeness.

(2a) **Soundness**: for every $f \notin L_{1,j-1}\backslash L_{1,j}$, $f$ will not be inserted to $\mathsf{inst}^{I:\Delta^-}_p$.

There are two cases:
(i) $f \notin L_{1,j-1}$ and $f \notin \Pi_p[I^r_{1,j}, \Delta^r_{1,j}]$.
(ii) $f \in L_{1,j-1}$ and $f \in L_{1,j}$, therefore $f \in \Pi_p[I^r_{1,j}\backslash\Delta^r_{1,j}]$ but $f \notin \Pi_p[I^r_{1,j}, \Delta^r_{1,j}]$.
Both cases consist of $f \notin \Pi_p[I^r_{1,j}, \Delta^r_{1,j}]$, therefore, there is no way $f$ will be inserted to $\mathsf{inst}^{I:\Delta^-}_p$ in line 3.

(2b) **Completeness**: for every $f = \chi^r(p)\sigma$ that $f \in L_{1,j-1}$ but $f \notin L_{1,j}$, $f$ can be derived by line 3 and inserted to $\mathsf{inst}^{I:\Delta^-}_p$.

Since $f \in L_{1,j-1}\backslash L_{1,j}$, then $f$ must be in $L_{1,j-1}$ (i.e., $f \in \mathsf{inst}^I_p$) and
(i) $f \notin \Pi_p[I^r_{1,j}\backslash\Delta^r_{1,j}]$, or
(ii) $f \in \Pi_p[I^r_{1,j}\backslash\Delta^r_{1,j}]$, $f \in \Pi_p[I^r_{1,j}, \Delta^r_{1,j}]$.

In both cases, $f$ will be inserted to $\mathsf{inst}^{I:\Delta^-}_p$. The proof will be similar to the completeness of claim 3. $\square$

Then, we prove the $\mathsf{Red}^r$ call next (i.e., $C^r_{1,d_1+1}$) can update $\mathsf{inst}^r_p$ correctly.

**Claim 4.** $C^r_{1,d_1+1}(I^r_{1,d_1}, \Delta^r_{1,d_1})$ can update $\mathsf{inst}^I_p$ correctly from $\Pi_p[I^r_{1,d_1}]\backslash\Pi_p[I^r_{1,d_1-1}, \Delta^r_{1,d_1-1}]$ to $\Pi_p[I^r_{1,d_1}]$.

First, following claim 3, we show before $C^r_{1,d_1+1}$, the values of $\mathsf{inst}^I_p$ satisfy the condition described in claim 4. After $C^r_{1,d_1}$, $\mathsf{inst}^I_p = \Pi_p[I^r_{1,d_1-1}\backslash\Delta^r_{1,d_1-1}]\backslash\Pi_p[I^r_{1,d_1-1}, \Delta^r_{1,d_1-1}]$, and $I^r_{1,d_1-1}\backslash\Delta^r_{1,d_1-1} = I^r_{1,d_1}$.

Oracle can accurately maintain the counting [Hu et al., 2018] of instantiations (i.e., $\mathsf{inst}^I_p$), therefore for $O(f)$ will return true if and only if $f \in \Pi_p[I^r_{1,d_1}]$. Then, we only need to show for every $f \in \Pi_p[I^r_{1,d_1-1}, \Delta^r_{1,d_1-1}]$ that needs to be checked, $f$ also belong to $\mathsf{inst}^{re}_p$. Since $\mathsf{inst}^{re}_p$ consists of all the overdeleted instantiations that are deleted during $C^r_{1,1}$ to $C^r_{1,d_i}$. For every $f = \chi^r(p)\sigma \in \Pi_p[I^r_{1,d_1-1}, \Delta^r_{1,d_1-1}]$, $f$ will be deleted in $C^r_{1,d_i}$ and inserted to $\mathsf{inst}^{re}_p$. $\square$

**Claim 5.** *In the first incremental run $Q^r_1$, a series of $\mathsf{Add}^r$ calls $C^r_{1,d_1+k+1}(I^r_{1,d_1+k}, \Delta^r_{1,d_1+k})$ (for $1 \leq k < h_1 - d_1$) will update $\mathsf{inst}^I_p$ correctly from $\Pi_p[I^r_{1,d_1+k}\backslash\Delta^r_{1,d_1+k}]$ to $\Pi_p[I^r_{1,d_1+k+1}]$.*

Before $C^r_{1,d_1+2}$, $\mathsf{inst}^I_p = \Pi_p[I^r_{1,d_1}]$ and $I^r_{1,d_1} = I^r_{1,d_1+1}\backslash\Delta^r_{1,d_1+1}$, which satisfies the condition described in claim 5.

One can prove every $\mathsf{Add}^r$ call $(C^r_{1,d_1+k+1}(I^r_{1,d_1+k}, \Delta^r_{1,d_1+k})$ for $1 \leq k < h_1 - d_1$) updates $\mathsf{inst}^I_p$ correctly just like claim 1. $\square$

Combining claim 2, 3, 4, and 5, we can have the following claim:

**Claim 6.** *Every call in run $Q^r_1$ can update $\mathsf{inst}^I_p$ correctly as defined in claim 2, 3, 4, and 5.*

Also, just like claim 6, we can have the following claim:

**Claim 7.** *Every call in run $Q^r_i$ for $i \geq 1$ can update $\mathsf{inst}^I_p$ correctly. Specifically:*

- *the first $\mathsf{Del}^r$ call $C^r_{i,1}(I^r_{i,0}, \Delta^r_{i,0})$ will update $\mathsf{inst}^I_p$ correctly from $\Pi_p[I^r_{i,0}]$ to $\Pi_p[I^r_{i,0}\backslash\Delta^r_{i,0}]\backslash\Pi_p[I^r_{i,0}, \Delta^r_{i,0}]$;*

- *the following $\mathsf{Del}^r$ calls $C^r_{i,j+1}(I^r_{i,j}, \Delta^r_{i,j})$ for $1 \leq j < d_i$ can update $\mathsf{inst}^I_p$ from $\Pi_p[I^r_{i,j}]\backslash\Pi_p[I^r_{i,j-1}, \Delta^r_{i,j-1}]$ to $\Pi_p[I^r_{i,j}\backslash\Delta^r_{i,j}]\backslash\Pi_p[I^r_{i,j}, \Delta^r_{i,j}]$;*

- *Then the next $\mathsf{Red}^r$ call $C^r_{i,d_i+1}$ will update $\mathsf{inst}^I_p$ from $\Pi_p[I^r_{i,d_i}]\backslash\Pi_p[I^r_{i,d_i-1}, \Delta^r_{i,d_i-1}]$ to $\Pi_p[I^r_{i,d_i}]$;*

- *Finally, the $\mathsf{Add}^r$ calls $C^r_{i,d_i+k+1}(I^r_{i,d_i+k}, \Delta^r_{i,d_i+k})$ for $1 \leq k < h_i - d_i$ will update $\mathsf{inst}^I_p$ from $\Pi_p[I^r_{i,d_i+k}\backslash\Delta^r_{i,d_i+k}]$ to $\Pi_p[I^r_{i,d_i+k}]$.*

1. *Induction Base*: every step in $Q^r_0$ and $Q^r_1$ can correctly update $\mathsf{inst}^I_p$ just as described above (claim 1 and 6).

2. *Induction Step*: for every call $C^r_{i,j}$ for $1 \leq j \leq h_i$, we can prove it updates $\mathsf{inst}^I_p$ correctly as described above. The proof will be just like claim 2, 3, 4, and 5. $\square$

Besides, we also prove that after the domain of each node $p$ is fixed, the improved Yannakakis algorithm ( Line 6 to 8 of algorithm 3) can correctly join the data that is currently active in each node.

**Claim 8.** *Line 6 to 8 of algorithm 3 can correctly join $\mathsf{inst}^{ac}_p$ for each $p \in N^r$ and project to $\mathsf{var}(\mathsf{h}(r))$, i.e., computing $\pi_{\mathsf{var}(\mathsf{h}(r))}(\bowtie_{p \in N^r} \mathsf{inst}^{ac}_p)$.*

Line 7 to 8 are steps of standard Yannakakis algorithms [Yannakakis, 1981], therefore, using line 7 and 8 can correctly compute the data described in claim 8. Then, if we could prove the values of $\mathsf{inst}^{ac}_p$ after using line 6 and 7 to reduce data are exactly the same as $\mathsf{inst}^{ac}_p$ when using only 7

to reduce data. In our words, the left semi-join sequence we add in line 6 won't influence the results of the full reducer.

For a node $p$ in the tree, Bernstein and Chiu [1981] defined a sequence of left semi-joins for $p$ (i.e., $\mathsf{LSJ}_p$) and proved that as long as $\mathsf{LSJ}_p$ is embedded in $\mathsf{LSJ}$, then $\mathsf{LSJ}$ can fully reduce the data in $p$. Additionally, they proved that the shortest left semi-join sequence that can fully reduce all the nodes for a tree-like query has the length of $2 \cdot (n-1)$, and the sequence is exactly what we present in line 7. Therefore, by adding another left semi-join sequence (line 6) before line 7, the valid $\mathsf{LSJ}_p$ that can fully reduce $p$ for every $p \in N^r$ is still included in the final used left semi-join sequence. So, the results of using line 6 and 7 are the same as the results of using only line 6. □

## D Proof of Lemma 1

**Lemma 1**: Algorithnm 2 computes $r[I \vdots \Delta^+]\backslash I$.

**Proof of Lemma 1**. In our paper, we use the complete hypertree decomposition $T^r$ for every $r$, i.e., every $B_i$ in $\mathsf{b}(r)$ is strongly covered. In other words, for every body atom $B_i \in \mathsf{b}(r)$, there exist at least one node $p$ s.t. $B_i \in \lambda^r(p)$ and $\mathsf{var}(B_i) \subseteq \chi^r(p)$. We prove Lemma 1 by showing its soundness and completeness.

1. **Soundness**: For every $f \notin r[I \vdots \Delta^+]\backslash I$, $f$ will not be returned by $\mathsf{Add}^r$.

   (1a) if $f \in r[I \vdots \Delta^+]$ and $f \in I$, then $f$ will be removed in line 10.

   (1b) if $f \notin r[I \vdots \Delta^+]$, then $\mathsf{b}(r\sigma) \not\subseteq I$ or $\mathsf{b}(r\sigma) \subseteq I$ but $\mathsf{b}(r\sigma) \cap \Delta^+ = \emptyset$. For the former case, such a substitution $\sigma$ is not a valid substitution for operator $\Pi_p[I, \Delta^+]$, so instantiations that use this $\sigma$ will not be included in $\mathsf{inst}_p^I$ or $\mathsf{inst}_p^{I \vdots \Delta^+}$. Therefore, no data using this $\sigma$ will be derived. For the latter case, if $\mathsf{b}(r\sigma) \cap \Delta^+ = \emptyset$, then for all $B_i \in \mathsf{b}(r)$, $B_i\sigma \cap \Delta^+ = \emptyset$. Therefore, for every node $p \in N^r$, $\lambda^r(p)\sigma \cap \Delta^+ = \emptyset$. So the instantiations that use $\sigma$ are included in the $\mathsf{inst}_p^I$. In line 2 of algorithm 3, our algorithms force one node using the newly added instantiations by labeling this node as $\Delta^+$. So in this case, $f$ will not be derived.

2. **Completeness**: for every $f = \mathsf{h}(r\sigma)$ that $f \in r[I \vdots \Delta^+]\backslash I$, $f$ will be returned by $\mathsf{Add}^r$.

   Since $f \in r[I \vdots \Delta^+]\backslash I$, so we can have $\mathsf{h}(r\sigma) \notin I$, $\mathsf{b}(r\sigma) \subseteq I$, and $\mathsf{b}(r\sigma) \cap \Delta^+ \neq \emptyset$ for this $\sigma$. So there exists at least one $B_i$ s.t. $B_i\sigma \subseteq I$, and $B_i\sigma \cap \Delta^+ \neq \emptyset$. We define a set $P_{B_i,\sigma}^r$ as a subset of $N^r$ (the nodes in $T^r$) s.t. $B_i$ is assigned to these nodes and $\mathsf{inst}_p^{I \vdots \Delta^+}$ consists of new instantiations using $\sigma$. Formally:

$$P_{B_i,\sigma}^r = \{p \in N^r \mid B_i \in \lambda^r(p) \text{ and } \chi^r(p)\sigma \notin \mathsf{inst}_p^I\}. \quad (7)$$

For this substitution $\sigma$ that $f = \mathsf{h}(r\sigma)$, we define a set of nodes $P_\sigma^r$ whose instantiations $\mathsf{inst}_p^{I \vdots \Delta^+}$ is updated by this substitution $\sigma$. Formally:

$$P_\sigma^r = \{P_{B_i,\sigma}^r \text{ for every } B_i \in \mathsf{b}(r) \text{ if } B_i\sigma \cap \Delta^+ \neq \emptyset\} \quad (8)$$

Since $T^r$ is complete, there is at least one node in $P_\sigma^r$ for every $\sigma$ s.t. $f = \mathsf{h}(r\sigma) \in r[I, \Delta^+]\backslash I$. For node $p \in P_\sigma^r$, $\lambda^r(p)\sigma \cap \Delta^+ \neq \emptyset$ and $\chi^r(p)\sigma \notin \mathsf{inst}_p^I$, so the instantiation that uses $\sigma$ will be derived in line 3 by operator $\Pi_p[I, \Delta^+]$ and will be included in $\mathsf{inst}_p^{I \vdots \Delta^+}$.

For other nodes that $p' \notin P_\sigma^r$, the instantiation that uses $\sigma$ is included in the current $\mathsf{inst}_p^I$. This is because: assume there is a $p' \notin P_\sigma^r$ and $\chi^r(p')\sigma \notin \mathsf{inst}_p^I$. Then (1) if there exists $B_i$ s.t. $B_i \in \lambda^r(p')$ and $B_i\sigma \cap \Delta^+ \neq \emptyset$, then according to expression (7), $p' \in P_{B_i,\sigma}^r \subseteq P_\sigma^r$, which is contradict to our assumption; (2) if no such $B_i$ exists, then $\lambda^r(p')\sigma \cap \Delta^+ = \emptyset$, i.e., $\lambda^r(p')\sigma \subseteq I\backslash \Delta^+$, then according to claim 1 and 7, $\chi^r(p')\sigma \in \mathsf{inst}_p^I = \Pi_p[I\backslash \Delta^+]$.

Let $\max(P_\sigma^r)$ be the nodes in $P_\sigma^r$ whose index is the maximum in the prefixed order of all the nodes in $N^r$. When $\max(P_\sigma^r)$ is labelled as $\Delta^+$, therefore the domain of $\mathsf{inst}_p^{I \vdots \Delta^+}$ is used; while any other nodes in $P_\sigma^r$ are labelled as $I \cup \Delta^+$, so the domain of $\mathsf{inst}_p^I \cup \mathsf{inst}_p^{I \vdots \Delta^+}$ is used. In both domains, the instantiations that use $\sigma$ are included. For $p' \notin P_\sigma^r$, its label will be $I$ or $I \cup \Delta^+$, either way, the instantiations that use $\sigma$ are included. Because line 6 to 8 of algorithm 3 can correctly join data between nodes (claim 8), $f$ can be derived in line 5 of algorithm 2 and returned by $\mathsf{Add}^r$. □

## E Proof of Lemma 2

**Lemma 2**: Algorithm 4 computes $r[I \vdots \Delta^-] \cap (I\backslash \Delta^-)$.

**Proof of Lemma 2**. Just like the proof of lemma 1, we can prove lemma 2 by showing its soundness and completeness.

1. **Soundness**: for every $f \notin r[I \vdots \Delta^-] \cap (I\backslash \Delta^-)$, $f$ will not be returned by $\mathsf{Del}^r$.

   (1a) $f \in r[I \vdots \Delta^-]$ but $f \notin I\backslash \Delta^-$. Such a $f$ will be removed in line 11 of algorithm 4.

   (1b) $f \notin r[I \vdots \Delta^-]$. This case is similar to the (1b) of the proof of lemma 1.

2. **Completeness**: For every $f \in r[I \vdots \Delta^-] \cap (I\backslash \Delta^-)$, $f$ will be returned by $\mathsf{Del}^r$.

   The proof of completeness is similar to lemma 1. □

## F Proof of Lemma 3

**Lemma 3**: Algorithm 5 computes $r[I] \cap \Delta$.

**Proof of Lemma 3**. We first define $\mathsf{inst}_p^\sigma$ as a single instantiation that is instantiated by the substitution $\sigma$, in which $\sigma$ is defined on $\mathsf{var}(\mathsf{b}(r))$. Formally:

$$\mathsf{inst}_p^\sigma = \chi^r(p)\sigma. \quad (9)$$

For a tuple $f = \mathsf{h}(r\sigma)$ in the database, the Oracle will maintain the counting of different valid substitutions to derive $f$. Two substitutions $\sigma, \sigma'$ that can derive the same $f$ are different if there exists at least a node $p \in N^r$ s.t. $\mathsf{inst}_p^\sigma \neq \mathsf{inst}_p^{\sigma'}$. For a tuple $f = \mathsf{h}(r\sigma)$ in the database, $\mathsf{Oracle}(f)$ will return true if there exists at least one remaining $\sigma$ defined on

var(b($r$)) s.t. $f$ can be derived by joining $\text{inst}_p^\sigma$ for each $p \in N^r$ and $\text{inst}_p^\sigma \in \text{inst}_p^I$ (we refer the values of $\text{inst}_p^I$ before line 5 of algorithm 5, same below).

For an instantiation $f_p \in \text{inst}_p^I$, the Oracle will maintain the counting of different valid substitutions to derive $f_p$. Two different substitutions $\sigma_p, \sigma_p'$ that can derive the same $f_p$ are different if $\lambda^r(p)\sigma_p \neq \lambda^r(p)\sigma_p'$. Actually, the counting maintained by Oracle is equivalent to the recursive counting introduced by Hu *et al.* [2018] and they also proved the correctness of recursive counting. Each node $p \in N^r$ can also be regarded as a subrule of the original rule $r$. Therefore, in some cases, the Oracle only maintains the counting of "half" of the step of rule $r$. So after rederiving the instantiations in $p$, we need to evaluate the tree to rederive the data introduced by the original rule $r$.

Formally, for every $f = \mathsf{h}(r\sigma) \in r[I] \cap \Delta$, it will have two cases: (1) there exists a substitution $\sigma$ s.t. $\text{inst}_p^\sigma \in \text{inst}_p^I$ (using the values of $\text{inst}_p^I$ before line 5). In this case, $f$ can be safely rederived by using Oracle in line 7 of algorithm 5. (2) in the other case, there is at least one node $p' \in N^r$ that $\text{inst}_{p'}^\sigma \notin \text{inst}_{p'}^I$, so $f$ can't be derived by (1) using Oracle directly. However, as we proved in claim 7, in-node rederivation (line 2) can accurately update $\text{inst}_{p'}^{I \vdots \Delta^+}$ to $\Pi_{p'}[I] \setminus \text{inst}_{p'}^I$ and update $\text{inst}_{p'}^I$ to $\Pi_{p'}[I]$ in line 5. Since $f \in r[I]$, we have $\mathsf{b}(r\sigma) \subseteq I$, and thus $\lambda^r(p')\sigma \subseteq \Pi_{p'}[I]$. Therefore, after in-node rederivation, $\text{inst}_{p'}^\sigma \in \text{inst}_{p'}^{I \vdots \Delta^+} \subseteq \Pi_{p'}[I]$. So we can also derive $f$ in this case.

The discussion above is about completeness. For the soundness of algorithm 5, since we only check tuples in $\Delta$ that are overdeleted, $\Delta$ strictly constrains the domain of the data that can be rederived. Besides, the soundness of Oracle has been proved by Hu *et al.* [2018]. Therefore, algorithm 5 is sound. □

## References

Philip A Bernstein and Dah-Ming W Chiu. Using semi-joins to solve relational queries. *Journal of the ACM (JACM)*, 28(1):25–40, 1981.

Pan Hu, Boris Motik, and Ian Horrocks. Optimised maintenance of datalog materialisations. In *Proceedings of the AAAI Conference on Artificial Intelligence*, volume 32, 2018.

Pan Hu, Boris Motik, and Ian Horrocks. Modular materialisation of datalog programs. *Artificial Intelligence*, 308:103726, 2022.

Mihalis Yannakakis. Algorithms for acyclic database schemes. In *VLDB*, volume 81, pages 82–94, 1981.



\end{document}